\documentclass[prd,preprint,showpacs,showkeys,
superscriptaddress,nofootinbib,floatfix]{revtex4}

\usepackage{graphicx}


\begin{document}

\setcounter{page}{0}
\thispagestyle{empty}

\preprint{$
\begin{array}{l}
\mbox{ANL-HEP-PR-04-59}\\
\mbox{CTEQ-0464}\\
\mbox{hep-ph/0406143}\\[5mm]
\end{array}
$}
\vspace*{-0.2cm}

\title{Light Gluino Constituents of Hadrons \\ and a 
Global Analysis of Hadron Scattering Data}

\author{Edmond~L.~Berger}
\email[e-mail: ]{berger@anl.gov}
\affiliation{High Energy Physics Division, Argonne National Laboratory,
Argonne, IL 60439}
\author{Pavel M. Nadolsky}
\email[e-mail: ]{nadolsky@mail.physics.smu.edu}
\affiliation{Department of Physics, Southern Methodist University, Dallas,
TX 75275-0175}
\author{Fredrick~I.~Olness}
\email[e-mail: ]{olness@smu.edu}
\affiliation{Department of Physics, Southern Methodist University, Dallas,
TX 75275-0175}
\author{Jon Pumplin}
\email[e-mail: ]{pumplin@pa.msu.edu}
\affiliation{Department of Physics \& Astronomy, Michigan State University,
East Lansing, MI 48824}

\date{\today}

%



\begin{abstract}
\noindent Light strongly interacting supersymmetric particles may be 
treated as partonic constituents of nucleons in high energy scattering 
processes.  We construct parton distribution functions for protons in 
which a light gluino is included along with standard model quark, antiquark, 
and gluon constituents.  A global analysis is performed of a large set 
of data from deep-inelastic lepton scattering, massive lepton pair 
and vector boson
production, and hadron jet production at large values of transverse momentum.  
Constraints are obtained on the allowed range of gluino mass as a function of 
the value of the strong coupling strength $\alpha_s(M_Z)$ determined at the 
scale of the $Z$ boson mass.  We find that gluino masses as small as $10$~GeV 
are admissible provided that $\alpha_s(M_Z) \ge 0.12$.  Current hadron 
scattering data are insensitive to the presence of gluinos heavier than 
$\sim 100 - 150$~GeV.     
\end{abstract}



\maketitle

\newpage

\section{Introduction\label{sec:Introduction}}

Relatively light strongly-interacting fundamental particles may 
be considered as constituents of nucleons.  The nature of these 
constituents and their experimental effects become evident when 
the parent hadrons are probed at sufficiently short distances 
or, equivalently, sufficiently large four-momentum 
transfer $Q$.  The charm quark $q = c$ and the bottom 
quark $b$ are treated appropriately as constituents of hadrons in 
situations in which $Q > m_q$, where $m_q$ is the mass of the heavy 
quark.  Other strongly interacting fundamental particles may exist, 
as yet undiscovered experimentally, with masses lying somewhere 
between the bottom- and top-quark masses.  One example is a relatively 
light gluino: a color-octet fermion and the supersymmetric partner 
of the massless spin-1 gluon.  For our purposes, we define 
a ``light'' particle to have a mass less than 100 GeV. 
In this paper, we explore the effects 
that a color-octet fermion would have on the parton 
distribution functions of nucleons, with a view toward establishing 
whether the set of hard-scattering data used in global analysis may 
already place significant constraints on the existence and allowed 
masses of such states.           

In our investigation, we use a light gluino from 
supersymmetry (SUSY)~\cite{Nilles:1984ge,Haber:1985rc} 
as a concrete example, but our analysis and conclusions should apply 
as well to the case of a color-octet fermion of whatever origin.  
As constituents of hadrons, these color-octet fermions share the 
momentum of the parent hadron with their standard model quark, 
antiquark, and gluon partners.  The distribution of light-cone momentum 
fraction $x$ carried by constituents is specified by parton distribution 
functions (PDFs) as functions of both $x$ and the scale $Q$ of the 
short-distance hard scattering.  The process-independent PDFs are 
essential ingredients for obtaining normalized predictions of rates for  
hard-scattering reactions at high energies.  
A simultaneous analysis of a large body of scattering data (global 
analysis) provides strong constraints on the magnitude and $x$ dependence 
of the PDFs.  

In perturbative quantum chromodynamics (QCD), the existence of a color-octet 
fermion and its couplings to the standard model constituents alter the coupled  
set of evolution equations that governs the functional change of the parton 
distributions as momentum is varied.  Gluinos have different renormalization 
group properties from those of the quark and gluon constituents, and  
the contributions of fermions in the color-octet representation  
are enhanced strongly.  Within the context of broken supersymmetry, squarks 
(scalar partners of quarks) may also be relatively light, particularly 
those of the third generation, the bottom- and 
top-squarks~\cite{Martin:1997ns,Dawson:1997tz,Dedes:2000nv,Berger:2000mp}.  
In the study reported here, we include a gluino in our PDF analysis, 
but we neglect possible 
contributions from other hypothesized supersymmetric states with 
masses above a few GeV, such as bottom squarks. 
As explained in Sec.~II, the effects of squark contributions 
on the current data are much less 
important than those of gluinos.  The approximation of retaining only the 
light gluino contribution simplifies the calculations while retaining most 
of the relevant physics.  

In a global analysis of hadronic data, a large sample of data is studied 
(about 2000 points) from a variety of experiments at different momentum scales.
The data set included in our study is the same as in the recent 
CTEQ6~\cite{Pumplin:2002vw} study done within the context of the standard 
model.  The data come from deep-inelastic lepton scattering, 
massive lepton pair 
and gauge boson production, and hadron jet production at large values of 
transverse momentum.  We apply the methodology of the next-to-leading order 
(NLO) CTEQ6 analysis to explore the compatibility of a light gluino with the 
large set of hadronic data.  Methods developed recently for the analysis of 
uncertainties of 
PDFs~\cite{Giele:1998gw,Alekhin:2000ch,Giele:2001mr,Botje:1999dj,Barone:1999yv,Pumplin:2000vx,Pumplin:2001ct,Stump:2001gu}
allow us to obtain quantitative bounds on the existence and masses of 
gluinos from a global analysis.  In early PDF analyses within the context of 
light gluinos~\cite{Roberts:1993gd,Blumlein:1994kw,Ruckl:1994bh}, a
gluino with a mass 5 GeV or less 
was found to be consistent with the data available at that time. 
A more recent study~\cite{Li:1998vr} disfavors a gluino with a mass
1.6 GeV or less. The much larger 
sample of the data in the modern fit and improved understanding of PDF 
uncertainties make it possible to derive more precise bounds. 

Light superpartners influence the evolution with scale $Q$ of the strong 
coupling strength $\alpha_s(Q)$.  The constraints we obtain on the gluino mass 
from a global 
analysis depend significantly on the value of the strong coupling strength 
$\alpha_s(M_Z)$ that is an ingredient in the global analysis.  In Sec.~II, we 
begin with a brief review of the dominant experimental constraints on $\alpha_s$ 
and consider the changes that may arise if supersymmetric particles and 
processes are present.  Further discussion of experimental constraints on 
$\alpha_s(M_Z)$ may be found in the Appendix.  We describe in Sec.~II.B how we 
implement the NLO evolution of 
the PDFs, while including the gluino degree of freedom at leading order (LO).  
Once supersymmetric particles are admitted, they contribute to hard scattering 
processes either as incident partons and/or as produced particles.  We 
therefore specify the hard scattering matrix elements that describe supersymmetric 
contributions to the rate for jet production at large transverse 
momentum.  In Sec.~III, we present and discuss the results of our global fits.  
The discriminating power of our analysis depends crucially on the inclusion of 
the Tevatron jet data in the fit. The inclusion of a light gluino in the PDFs 
removes momentum from the gluon PDF at large $x$, tending to depress the 
contribution from SM processes to the jet rate at large $E_T$.  However, the 
effect is compensated partially by a larger value of $\alpha_s(M_Z)$, slower 
evolution of $\alpha_s$ that makes $\alpha_s(E_T > M_Z)$ larger than in the 
standard model, and by contributions to the jet rate from production of SUSY 
particles in the final state.  

Our conclusions are summarized in Sec.~IV.  We find that the hadron scattering 
data provide significant constraints on the existence of gluinos whose mass is 
less than the weak scale $\sim 100$~GeV.  A large region of gluino parameter 
space is excluded by the global analysis independently of direct 
searches or other indirect methods.  The quantitative lower bounds we 
obtain on the gluino mass must be stated in terms of the assumed value of the 
the strong coupling strength $\alpha_s(M_Z)$.  For the standard model 
world-average value $\alpha _{s}(M_{Z})=0.118$, gluinos lighter than $12$~GeV 
are disfavored.  However, the lower bound on $m_{\tilde{g}}$ is relaxed to 
less than $10$~GeV if $\alpha _{s}(M_{Z})$ is increased above $0.120$.

\section{$\alpha_s$, Parton Densities, and Hard-Scattering Subprocesses}\label{sec:alphaS}

The presence of a light gluino $\tilde{g}$ and/or a light 
squark $\tilde{q}$ modifies the PDF global analysis in three ways.  
First, the gluino and squark change the evolution of the 
strong coupling strength $\alpha_s(Q)$ as the scale $Q$ is varied.
Second, the gluino and squark provide additional partonic degrees of freedom
that share in the nucleon's momentum and affect the PDFs of the standard 
model partons, e.g., via the channels
$g\rightarrow \tilde{g}\tilde{g}$ and 
$q\rightarrow \tilde{q}\tilde{g}$.
Third, gluino and squark contributions play a role in the hard-scattering 
matrix elements for the physical processes for which data are analyzed 
and fitted.  We discuss each of these modifications in the following three 
subsections.

\subsection{Modified evolution and values of $\alpha_s(Q)$}

The expansion of the evolution equation for $\alpha_s(Q)$ as a power series 
in $\alpha_s(Q)$ is  
\begin{eqnarray}
Q \frac{\partial }{\partial Q }\, \, \alpha_{s}(Q ) & = & -\frac{\alpha_{s}^{2}}{2\pi }\sum _{n=0}^{\infty }\beta _{n}\left(\frac{\alpha_{s}}{4\pi }\right)^{n}\nonumber \\
 & = & -\left[\beta _{0}\frac{\alpha_{s}^{2}}{2\pi }+\beta _{1}\frac{\alpha_{s}^{3}}{2^{3}\pi ^{2}}+...\right].\label{AlphaSEvolution}
\end{eqnarray}
When supersymmetric particles are included, the first two
coefficients in Eq.~(\ref{AlphaSEvolution}) are 
(see, e.g., Ref.~\cite{Machacek:1983tz})\begin{eqnarray}
\beta _{0} & = & 11-\frac{2}{3}n_{f}-2n_{\tilde{g}}-\frac{1}{6}n_{\tilde{f}} ,\label{beta0}
\end{eqnarray}
and \begin{eqnarray}
\beta _{1} & = & 102-\frac{38}{3}n_{f}-48n_{\tilde{g}}-\frac{11}{3}n_{\tilde{f}}+\frac{13}{3}n_{\tilde{g}}n_{\tilde{f}} ,\label{beta1}
\end{eqnarray}
where $n_{f}$ is the number of quark flavors, $n_{\tilde{g}}$
is the number of gluinos, and $n_{\tilde{f}}$ is the number of squark
flavors.  Equation~(\ref{beta0}) shows that, to the leading order, one 
generation of gluinos $\tilde{g}$ contributes the equivalent of 3 quark 
flavors to the QCD $\beta$-function. The effect of one squark flavor is 
equivalent to one-fourth of the contribution of a quark flavor.
In our work, we henceforth neglect the possibility of a light squark 
contribution to the $\beta$-function and limit ourselves to the effects of 
a light gluino. Inclusion of a light bottom squark changes the running of
$\alpha_s$ slightly, compatible with current 
data~\cite{Becher:2001zb,Becher:2002ue}. The modified coefficients $\beta _{0}$ 
and $\beta _{1}$ for
$n_{\tilde{g}}=1$ and $n_{\tilde{f}}=0$ are implemented
in our numerical calculation to full NLO accuracy.

In our global fit of hadron scattering data, the allowed range of the 
gluino mass $m_{\tilde{g}}$ depends strongly on the assumed value of the
strong coupling $\alpha_s(M_Z)$. Therefore, it is important to understand
the current experimental constraints on $\alpha _s(M_Z)$ from sources other 
than hadron scattering data.  A combined analysis of all $Z$-pole data within 
the context of the standard model, carried out by a working group of members 
of the four CERN Large Electron-Positron Collider (LEP) experiments and the 
SLAC SLD experiment~\cite{unknown:2003ih}, 
obtains the value 
$\alpha_s(M_Z) = 0.1187 \pm 0.0027$.\footnote{See Table 16.2 of 
Ref.~\cite{unknown:2003ih}.}  This value is but a shade greater than the 
often-quoted standard model world-average value 
$\alpha_s(M_Z) = 0.1183 \pm 0.0027$~\cite{Bethke:2002rv} obtained 
from a variety of determinations of $\alpha_s(Q)$ 
at different momentum scales $Q$.  On the other hand, the value of 
$\alpha_s(Q)$ extracted from data in the context of supersymmetric 
contributions can be different from the value obtained in standard model 
fits.  Some of the assumptions made in a standard model analysis are 
modified by the presence of the supersymmetric 
contributions~\cite{Berger:2000mp,Berger:2001jb,Chiang:2002wi,Luo:2003uw}, 
and the value of $\alpha_s(Q )$ at $Q =M_{Z}$ obtained in standard 
model analyses may have to be be revised.  A recent 
estimate~\cite{Luo:2003uw} of SUSY effects on directly measured $\alpha_s(M_{Z})$
finds values in the interval $(0.118-0.126)\pm 0.005$, where the
variation in the central value arises from uncertainty in the magnitude
of SUSY-QCD corrections to the $Z$-boson decay width from processes such 
as $Z \to b\bar{\tilde b}{\tilde g}$ and 
$Z \to {\bar b}{\tilde b}{\tilde g}$.  Further discussion of the evolution 
of $\alpha_s(Q)$ in the presence of light 
supersymmetric states may be found in the Appendix. 

In a general purpose CTEQ fit, $\alpha_{s}(M_{Z})$ is fixed to be
equal to its world-average value, determined from a combination of the 
$\tau$-lepton decay rate, LEP $Z$ pole observables, and other measurements. 
As discussed above,
this value may change in the presence of light superpartners. To explore
fully the range of strong coupling strengths compatible with the global
fit, we perform a series of fits in which $\alpha_{s}(M_{Z})$ is
varied over a wide range $0.110\leq\alpha_{s}(M_{Z})\leq0.150$. We
then determine the values of $m_{\tilde{g}}$ and $\alpha_{s}(M_{Z})$
preferred by the global fits.

\subsection{Implementation of a gluino in the NLO evolution of parton 
distributions\label{sec:DGLAP}}

In the construction of parton distributions, we include a light gluino 
and omit squark contributions.  A squark enters parton splitting functions 
only in combination with another rare particle, and these splittings are 
characterized by smaller color factors than in
the gluino case.  We incorporate the gluino sector into the PDF 
evolution package used to build the CTEQ6 unpolarized parton 
distributions~\cite{Pumplin:2002vw}.  

The standard procedure for extracting parton distribution functions from 
global QCD analysis is to parametrize the distributions at a fixed
small momentum scale $Q_0$.  The distributions at all higher $Q$ are
determined from these by the Dokshitzer-Gribov-Lipatov-Altarelli-Parisi
(DGLAP) evolution equations~\cite{Dokshitzer:1977sg,Gribov:1972ri,Altarelli:1977zs}.
The agreement with experiment is measured by an effective $\chi^{2}$, which
can be defined by $\chi^{2} =\sum_{\rm expts}\chi_{n}^{2}$, or by
generalizations
of that formula to include published systematic error correlations.
The PDF shape parameters at $Q_0$ are chosen to minimize $\chi^{2}$ and 
obtain the ``best fit'' PDFs.

We choose the starting value 
$Q_{0}$ for the QCD evolution equal to the smaller of the gluino mass 
$m_{\tilde{g}}$ or charm quark mass $m_{c}$.  At the scale $Q =Q_{0}$, 
the only non-perturbative input distributions are those of the gluons $g$ 
and light ($u,\, \, d,\, \, s$) quarks.  Non-zero PDFs of the gluinos 
and heavy quarks ($c,\, \, b$) are generated radiatively above their 
corresponding mass thresholds. In the CTEQ6 analysis, $Q_{0}$ coincides 
with the charm quark mass: $Q_{0}=m_{c}=1.3$ GeV. Therefore,
for $m_{\tilde{g}}\geq m_{c}$ the input scale $Q_{0}=1.3$~GeV is
the same as in the CTEQ6 study.  We use the CTEQ6 functional 
forms for the input PDFs of the standard model partons at $Q = Q_0$, 
but the starting values of the parameters are varied in order 
to obtain acceptable fits to the full set of scattering data.  

The prescription for $Q_{0}$ allows us to investigate the possibility
of gluinos lighter than charm quarks ($m_{\tilde{g}}<m_{c}$).
We include fits for gluino masses $0.7\leq m_{\tilde{g}}\leq 1.3$~GeV 
by choosing $Q_{0}=m_{\tilde{g}}$. Such super-light
gluinos may be generated both via perturbative and nonperturbative
mechanisms, and, in principle, an independent phenomenological parametrization
must be introduced for the gluino PDF to describe nonperturbative
contributions. Our prescription for the region $m_{\tilde{g}}<m_{c}$
provides a particular model for such an input gluino parametrization,
similar in its spirit to the dynamical parton distributions of the
GRV group \cite{Gluck:1998xa}, as well as the procedure used in 
earlier light gluino 
analyses \cite{Ruckl:1994bh} and \cite{Li:1998vr}.~\footnote{In 
Ref.~\cite{Li:1998vr}, the PDFs are obtained from the procedure
described here, but at LO and without inclusion of the jet production 
data, for $m_{\tilde{g}}=0.5$ and $1.6$~GeV.} 

In the presence of a light gluino, the DGLAP equations 
must be extended to account for the new processes.  The
coupled evolution equations take the form 
\begin{eqnarray}
 &  & Q ^{2}\frac{d}{dQ ^{2}}\left(\begin{array}{c}
 \Sigma (x,Q ^{2})\\
 g(x,Q ^{2})\\
 \tilde{g}(x,Q ^{2})\end{array}
\right)=\frac{\alpha _{s}(Q^{2})}{2\pi }\, \, \times \nonumber \\
 & \times  & \int _{x}^{1}\frac{dy}{y}\left(\begin{array}{ccc}
 P_{\Sigma \Sigma }\bigl (x/y\bigr ) & P_{\Sigma g}\bigl (x/y\bigr ) & P_{\Sigma \tilde{g}}\bigl (x/y\bigr )\\
 P_{g\Sigma }\bigl (x/y\bigr ) & P_{gg}\bigl (x/y\bigr ) & P_{g\tilde{g}}\bigl (x/y\bigr )\\
 P_{\tilde{g}\Sigma }\bigl (x/y\bigr ) & P_{\tilde{g}g}\bigl (x/y\bigr ) & P_{\tilde{g}\tilde{g}}\bigl (x/y\bigr )\end{array}
\right)\left(\begin{array}{c}
 \Sigma (y,Q ^{2})\\
 g(y,Q ^{2})\\
 \tilde{g}(y,Q ^{2})\end{array}
\right)\, ;\label{SAP}
\end{eqnarray}
\begin{eqnarray}
\Sigma (x,Q ^{2}) = \sum_{i=u,d,s,...} (q_i(x,Q ^{2}) + {\bar q_i}(x,Q ^{2})).  
\end{eqnarray}
Here $\Sigma (x,Q ^{2})$, $g(x,Q ^{2})$, and $\tilde{g}(x,Q ^{2})$
are the singlet quark, gluon, and gluino distributions, respectively; 
$q_i(x,Q ^{2})$ and ${\bar q_i}(x,Q ^{2})$ are the quark and antiquark 
distributions for flavor $i$.  
The splitting functions $P_{ij}(x)$ may be found in the literature
\cite{Antoniadis:1983qw}. 

\begin{figure}
\begin{center}\includegraphics[  width=1.0\textwidth,
  keepaspectratio]{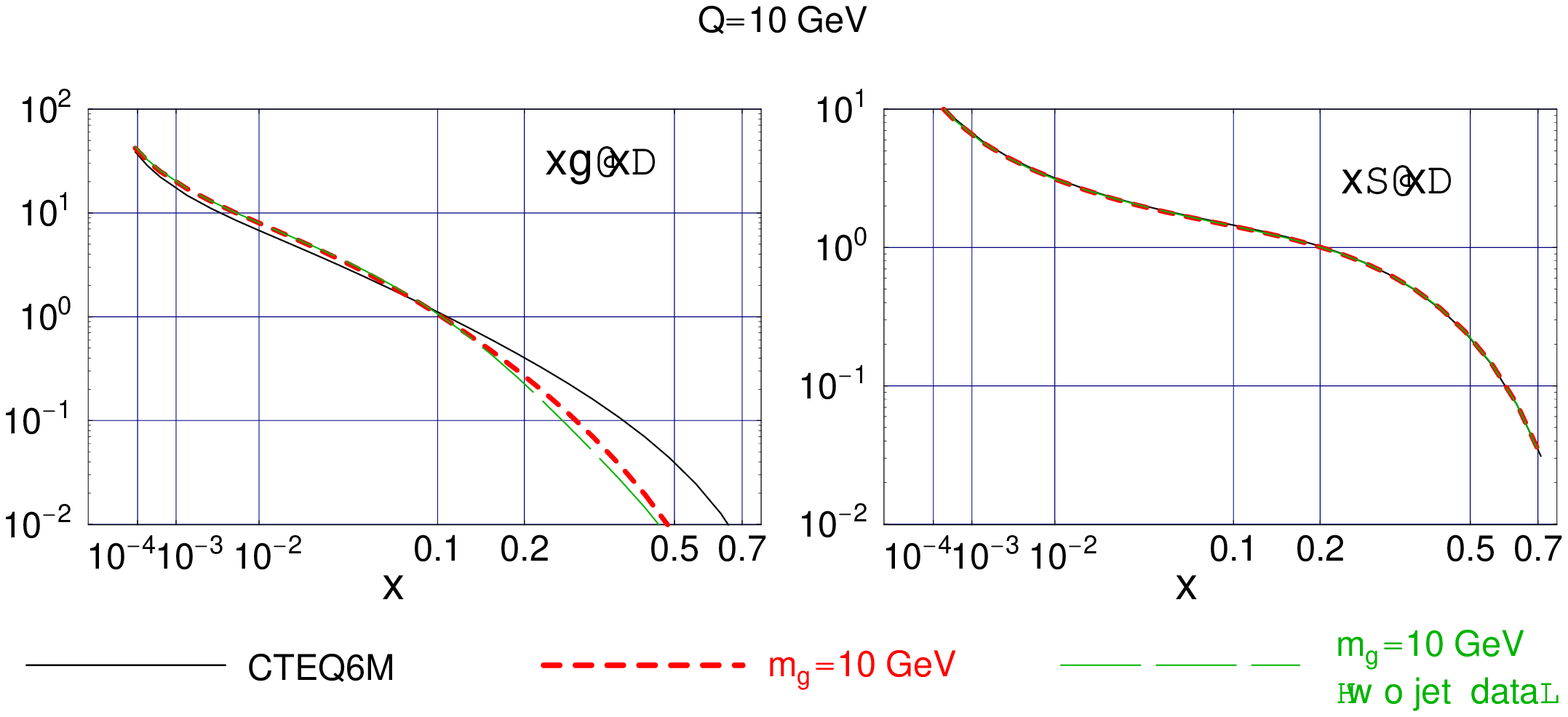}\end{center}

\caption{\label{fig:xf:q10} The gluon and singlet momentum distributions
$xg(x,Q^{2})$ and $x\Sigma (x,Q^{2})$ are displayed as functions of $x$ at 
$Q=10$~GeV with $\alpha _{s}(M_{Z})=0.118$ and $m_{\tilde{g}}=10$~GeV. The 
curves show the conventional CTEQ6M fit (solid) and the LG solutions with 
(short-dashed) and without (long-dashed) the Tevatron jet data included 
in the data set.}
\end{figure}

\begin{figure}
\begin{center}\includegraphics[  width=1.0\textwidth,
  keepaspectratio]{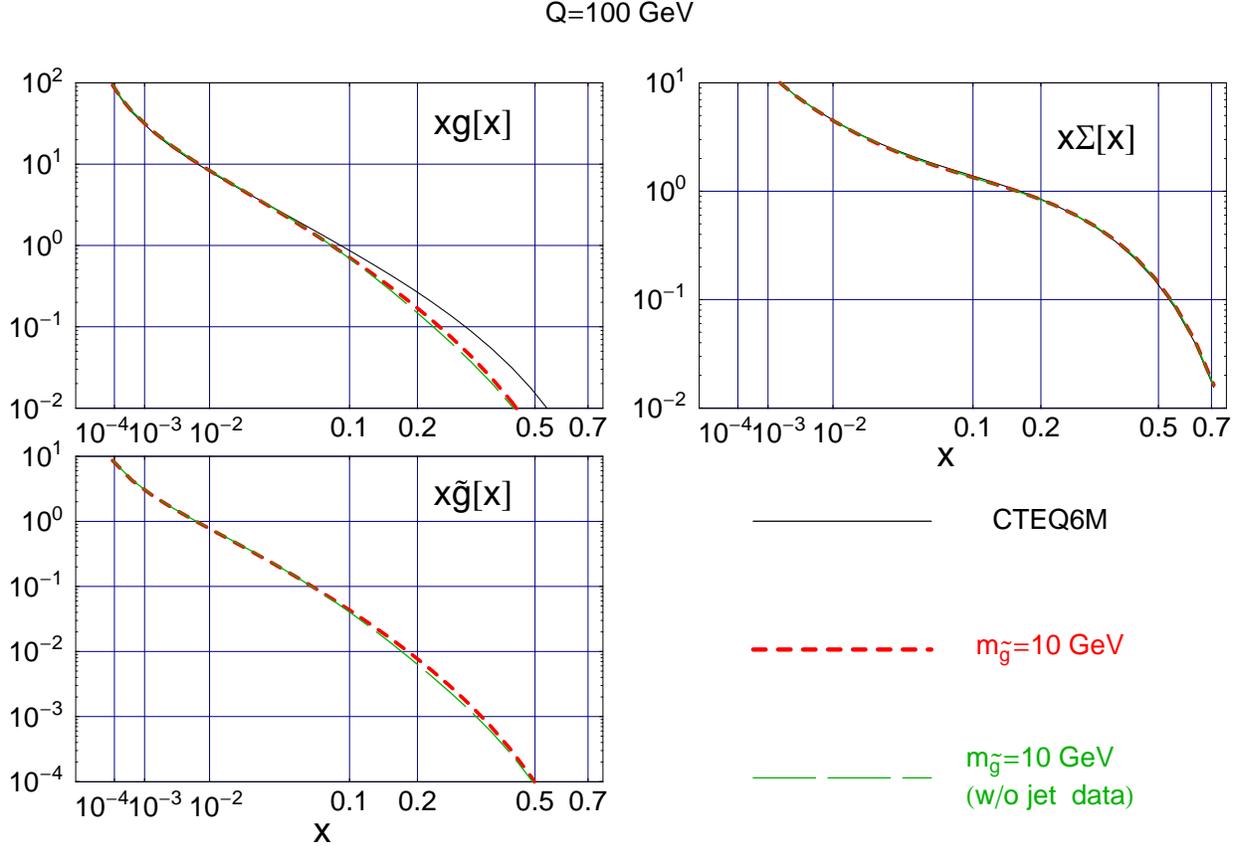}\end{center}

\caption{\label{fig:xf:q100} The gluon, singlet, and gluino momentum 
distributions $xg(x,Q^{2})$, $x\Sigma (x,Q^{2})$, and $x\tilde{g}(x,Q^{2})$ 
are displayed as functions of $x$ at $Q=100$~GeV with 
$\alpha _{s}(M_{Z})=0.118$ and $m_{\tilde{g}}=10$~GeV. 
The curves show the conventional CTEQ6M fit (solid) and the LG solutions 
with (short-dashed) and without (long-dashed) the Tevatron jet data included 
in the data set.}
\end{figure}

The inclusion of a gluino in the evolution equations complicates the 
calculation substantially. To achieve acceptable accuracy, evolution
of the quarks and gluons must certainly be done at next-to-leading
order accuracy. However, without a substantial loss in accuracy, we can 
simplify the overall calculation by evaluating the gluino contributions to 
leading order accuracy only. We use the following prescription:

\begin{enumerate}

\item Evolve the ordinary quarks and gluons at NLO (so that the splitting
functions $P_{\Sigma\Sigma},$ $P_{\Sigma g}$, $P_{g\Sigma}$, and $P_{gg}$ are evaluated
to order ${\mathcal{O}}(\alpha _{s}^{2})$). 

\item Evolve the gluinos at LO (so that the splitting functions $P_{g\tilde{g}},$
$P_{\Sigma \tilde{g}}$, $P_{\tilde{g}g}$, $P_{\tilde{g}\Sigma }$,
and $P_{\tilde{g}\tilde{g}}$ are evaluated to ${\mathcal{O}}(\alpha _{s})$).
In particular, at LO (and in the absence of the squarks),  
$P_{\tilde{g}\Sigma }=P_{\Sigma \tilde{g}}=0$. 

\item For the evolution of $\alpha_s$, use the full 
NLO (${\mathcal{O}}(\alpha _{s}^{2})$)
expression, \emph{including} the effect of the gluino.

\end{enumerate}
In this prescription, the evolution is fully accurate to NLO 
except for the gluino splitting kernels. 
Were we interested in a process dominated by gluino contributions,
we might need a NLO representation of the gluino PDF, 
$\tilde{g}(x,Q ^{2})$.
However, the impact of the gluino PDF is minimal for the inclusive 
data in the global analysis, since $\tilde{g}(x,Q ^{2})$
is much smaller than the quark and gluon PDFs 
(cf.~Figs.~\ref{fig:xf:q10} and \ref{fig:xf:q100}).  As a result, 
the gluino plays only an indirect role.  Its presence modifies the 
fit in two ways:

\begin{enumerate}
\item The gluino alters $\alpha_s(Q)$, thereby modifying the evolution
of ordinary quark and gluon PDFs. 
\item The gluino carries a finite fraction of the hadron's momentum, thereby
decreasing the momentum fraction available to the gluons and standard model 
quarks. 
\end{enumerate}

Regarding item (1), we compute the effects of the gluino correctly 
by using the exact NLO beta function that includes SUSY effects. Therefore, 
the only shortcoming of our prescription is with respect to item (2). We 
describe correctly the NLO mixing between the quarks and the gluons, 
but the less consequential mixing of 
the standard model partons and the gluino is correct only 
to leading order. In the energy range of our interest, 
the gluinos carry a small fraction ($\lesssim 5\%$) of the proton's momentum. 
The neglected NLO  corrections to this small quantity are
further suppressed by a factor of $\alpha_s/\pi$. They are
comparable in magnitude to the NNLO corrections for the standard model 
splittings, which are suppressed by $\alpha_s^2/\pi^2$.  
The uncertainty introduced by the omission of the NLO gluino
splittings is comparable to that due to the NNLO
corrections for other particles, and it may be ignored in
the present NLO analysis. 

Figures~\ref{fig:xf:q10} and~\ref{fig:xf:q100} illustrate and support 
our assumptions.  They show the momentum 
distributions for the gluons, singlet quarks, and 
gluinos, respectively.  In these figures, the strong coupling strength 
$\alpha _{s}(M_{Z})$ is equal to $0.118$ (the value of $\alpha _{s}(M_Z)$
assumed in the CTEQ6 analysis).  The momentum scale $Q$ is $10$~GeV in
Fig.~\ref{fig:xf:q10} and $100$~GeV in Fig.~\ref{fig:xf:q100}. 
The abscissa (x-axis) is scaled as $x^{1/3}$ in these plots of the 
dependence on the momentum fraction $x$.  The distributions are obtained 
from the light gluino (LG) fits for 
$m_{\tilde{g}}=10$~GeV, with or without
the inclusion of the Tevatron jet data.  For $xg(x,Q^{2})$
and $x\Sigma (x,Q^{2})$, we also show the corresponding distributions
from the best-fit set CTEQ6M of the CTEQ6 analysis.

With $Q=m_{\tilde{g}}=10$~GeV, the gluino density 
$x\tilde{g}(x,Q^{2}) =0$.  Nevertheless, the effects of gluino 
contributions on the fit at $Q>10$ GeV 
force a change in the gluon and quark distribution functions from 
their standard model values at $Q=m_{\tilde{g}}=10$~GeV, as shown 
in Fig.~\ref{fig:xf:q10}.  Once $Q$ is evolved to $100$~GeV, 
Fig.~\ref{fig:xf:q100} shows a nonzero momentum distribution for 
the gluinos and a persistent change of the gluon and quark densities 
from their CTEQ6M standard model values.  

The figures demonstrate two important features.  First, the magnitude 
of the gluino distribution is much smaller than the gluon and quark 
distributions.  This large difference justifies the assumptions 
that contributions are small 
from scattering subprocesses with initial-state gluinos, and that NLO 
gluino contributions may be omitted in our analysis.

Second, the presence of the gluino depletes the gluon
distribution at $x\gtrsim 0.05$.  The effect on the singlet distribution
is less pronounced.  The gluinos take their momentum
($3.7\%$ of the proton's momentum at $Q=100$~GeV 
for $m_{\tilde{g}}=10$~GeV) from the gluons
($3.0\%$) principally, less from quarks ($0.7\%$), independently
of whether the Tevatron jet data are included in the fit.  Since the
jet data at large transverse energy $E_T$ are known to probe the behavior 
of $g(x,Q^{2})$ at large $x,$ i.e., in the region where the depletion of 
the gluon's momentum is the strongest, we judge that inclusion of the jet 
data in the fit strengthens the constraining power of the fit. 

\subsection{Gluino contributions to hard scattering
\label{sec:HardMatrixElements}}

Once light superpartners are introduced as degrees of freedom, we
must consider their impact on all hard scattering processes. Their
effects can be felt both at tree level and in virtual-loop diagrams.
At leading order in perturbation theory, we may consider hard subprocesses
initiated by light gluinos or light bottom squarks that are constituents of the 
initial hadrons, as well as subprocesses in which gluinos or bottom
squarks 
are emitted in the final state.  We evaluate SUSY contributions
to the hard matrix elements at leading order only for the same reasons that 
justify the omission of NLO SUSY contributions to the splitting
kernels in Sec.~\ref{sec:DGLAP}.  


The CTEQ6 fit is performed to data from lepton-nucleon deep-inelastic
scattering (DIS), vector boson production (VBP), and hadronic jet
production at the Tevatron.  In DIS and VBP, the lowest-order contribution
from gluinos is $\gamma^* + g \rightarrow \tilde{q} + \tilde {g}$ at order 
${\mathcal{O}}(\alpha _{s})$. This subprocess proceeds via squark exchange, 
and its contribution can be neglected as being much smaller than the Born-level
QCD subprocesses (which contribute at order ${\mathcal{O}}(\alpha _{s}^{0})$).
The bottom squarks can contribute at ${\mathcal{O}}(\alpha _{s}^{0})$
through the subprocesses $\tilde{b}+\gamma ^{*}\rightarrow \tilde{b}$
in DIS and $\tilde{b}+\tilde{b}^* \rightarrow \gamma ^{*}$ in the 
Drell-Yan process. However, these contributions appear in a combination with
a small bottom squark PDF $\tilde{b}(x,Q ^{2})$ and, therefore, are also
negligible. We conclude that Born-level SUSY contributions appear only 
in the Tevatron jet production data, while, to the assumed level
of accuracy, the hard matrix elements in DIS and VBP remain the same
as in the SM case. 

We now consider the influence that gluino subprocesses may have on the rate 
for jet production at large values of transverse energy $E_{T}$.  Gluinos 
are color-octet fermions and, produced in the the final state, they 
materialize as jets.  Since the gluino parton density is relatively large 
only at small values of fractional momentum $x$ and, as illustrated in 
Figs.~\ref{fig:xf:q10} and \ref{fig:xf:q100}, is small even there when 
compared with the gluon and light-quark densities, we are justified 
in neglecting the contributions to the large $E_{T}$ jet rate from 
subprocesses initiated by two gluinos.  An example is 
$\tilde{g}+\tilde{g}\rightarrow g+g$.  However, in the 
interest of completeness, we include two subprocesses initiated by 
one gluino:  $g+\tilde{g}\rightarrow g+\tilde{g}$, and
$q+\tilde{g}\rightarrow q+\tilde{g}$.   Subprocesses initiated by 
gluons and/or light quarks can be important.  We include 
$g+g\rightarrow \tilde{g}+\tilde{g}$ via
either a direct channel gluon or a cross channel gluino, and 
$q+\bar{q}\rightarrow \tilde{g}+\tilde{g}$
via a direct channel gluon.  We can ignore the $t$-channel exchange 
diagrams that contribute to $q+\bar{q}\rightarrow \tilde{g}+\tilde{g}$.  
With the possible exception of
the bottom squark, the masses of most squarks are so large that the
relevant $t$-channel amplitudes are negligible. In the case of bottom
squark exchange, the two initial-state partons would be bottom quarks,
with small parton densities. For similar reasons, we may also ignore
subprocesses such as $q+g\rightarrow \tilde{g}+\tilde{q}$.

At the Tevatron $p\bar{p}$ collider, the $q\bar{q}$ partonic luminosity
is relatively large in the region of large $E_{T}$, and one might
expect naively that the subprocess 
$q+\bar{q}\rightarrow \tilde{g}+\tilde{g}$
would increase the jet rate significantly. However, just as for the
direct channel gluon subprocess 
$q+\bar{q}\rightarrow q^{\prime }+\bar{q}^{\prime }$
in standard QCD, the squared matrix element for 
$q+\bar{q}\rightarrow \tilde{g}+\tilde{g}$
is relatively small.

In our treatment of jet production, we compute the matrix elements
for the SUSY-QCD subprocesses at leading order.
Working at large $E_{T} \gg m_{\tilde{g}}$, we neglect the gluino mass
in the calculation. We include these matrix elements as additional
contributions to the jet rate in the fitting program, adding them to those 
of the NLO standard model QCD processes {[}${\mathcal{O}}(\alpha _{s}^{3})${]}
to obtain constraints from the inclusive jet data. 

As indicated in the previous subsection, inclusion of a light gluino in the 
PDF set removes momentum from the gluon 
PDF at large $x$, tending to depress the contribution from SM processes to the 
jet rate at large $E_T$.  However, as we show, the effect is compensated 
partially by a larger value of $\alpha_s(M_Z)$, by slower evolution of 
$\alpha_s$ that makes $\alpha_s(E_T > M_Z)$ larger than in the standard 
model, and by contributions to the jet rate from production of SUSY particles 
in the final state.

\section{Presentation of the Global Fits 
\label{sec:NumericalAnalysis}}

Our global fits are made to the complete set of data used in the 
CTEQ6 analysis, for several fixed values of the gluino
mass.  At this stage of the analysis, we do not impose a value of 
$\alpha _{s}(M_{Z})$, preferring to determine a range of 
values directly from the global analysis of hadronic scattering data.  
We perform fits to the hadronic data with $\alpha _{s}(M_{Z})$ set equal 
to one of several 
selected values in the range $0.110\leq \alpha _{s}(M_{Z})\leq 0.150$.  
As discussed in Sec.~\ref{sec:HardMatrixElements}, jet production at 
the Tevatron is the only process among those we include that acquires
Born-level contributions from the light gluinos. Most of the figures 
presented are for fits to the full set of data.  To gauge the 
effect of the jet data, we also show results of additional fits 
that do not include these data.  The numbers of experimental points 
in the fits with (without) the Tevatron jet data are 1811 (1688).

\subsection{Contour plots of $\Delta \chi ^{2}$ vs $\alpha _{s}$ and $m_{\tilde{g}}$}

\begin{figure}
\begin{center}\includegraphics[  width=1.0\textwidth,
  keepaspectratio]{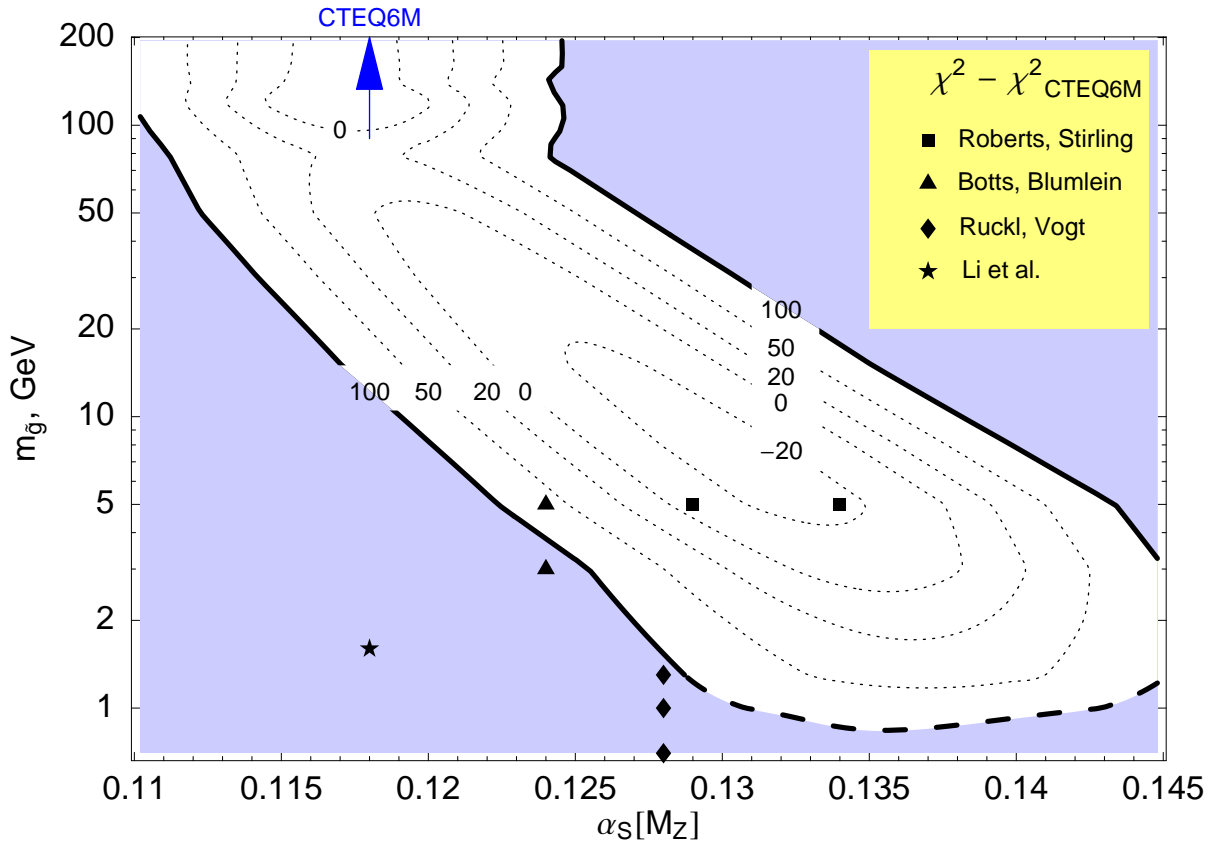}\end{center}

\caption{\label{fig:AllContours}A contour plot of 
$\Delta \chi ^{2}=\chi ^{2}\left(\alpha _{s}(M_{Z}),m_{\tilde{g}}\right)-\left.\chi ^{2}\right|_{\mbox {CTEQ6M}}$
as a function of the strong coupling $\alpha _{s}(M_{Z})$ and gluino
mass $m_{\tilde{g}}$. The values of $\Delta \chi ^{2}$ are shown
by labels on the corresponding isolines. The shaded region is excluded
by the CTEQ6 tolerance criterion. The points corresponding to the
earlier PDF fits with a 
LG~\cite{Roberts:1993gd,Blumlein:1994kw,Ruckl:1994bh,Li:1998vr}
are denoted by the symbols described in the legend. The solid line marks the 
$\Delta \chi ^{2}= 100$ isoline.  The dashed completion of this line at 
the bottom of the contour plot corresponds to fits done with 
$m_{\tilde{g}} = Q_0 \leq m_c = 1.3$~GeV.}
\end{figure}

The principal result of our analysis is shown in Fig.~\ref{fig:AllContours}.  
It maps the region of acceptable values of $\chi ^{2}$ in the plane
of $\alpha _{s}(M_{Z})$ and $m_{\tilde{g}}$.  The contour plot 
shows the difference 
$\Delta \chi ^{2}\equiv \chi ^{2}(\alpha _{s},m_{\tilde{g}})-\left.\chi ^{2}\right|_{\mbox {CTEQ6M}}$, 
between the value of $\chi ^{2}$ obtained in our LG fit and the standard model result 
equivalent to the CTEQ6M fit.  The point in the plane corresponding 
to the CTEQ6M fit ($\alpha _{s}(M_{Z})=0.118$ and $m_{\tilde{g}}\rightarrow \infty $)
is marked by the arrow.~\footnote{The best-fit value 
  $\alpha_s(M_Z)=0.117$ in the CTEQ6 fit is slightly below  
the world-average value $\alpha_s(M_Z)=0.118$ assumed in the CTEQ6M
PDF set.} The variation of $\chi^{2}$ in the neighborhood of the minimum 
is used to estimate limits of uncertainty.  
%
%

An overall tolerance parameter $T$ and a condition $\Delta \chi^2 < T^2$ 
are used in the CTEQ6 analysis to characterize the acceptable neighborhood 
around the global minimum of $\chi^2$ in the parton parameter space.  
The quantitative estimate $T=10$ is obtained from a combination
of the constraints placed on acceptable fits by each individual
experiment included in the fit~\cite{Pumplin:2002vw}.~\footnote
{The tolerance $T=10$ is estimated from the degree of consistency 
between the various data sets in the global fit.  It includes effects due 
to experimental uncertainty and uncertainties that are of theoretical or 
phenomenological origin. It is an oversimplification to represent all 
uncertainties of PDFs and their physical predictions by a single number 
$T$. However, given the complexity of the problem, it is unrealistic to 
be more precise at this stage. The criterion $T=10$ must be used with 
awareness of its limitations.}%

According to the tolerance on $\Delta \chi ^{2}$ of the CTEQ6 analysis, a fit is 
strongly disfavored if $\Delta \chi ^{2}>100$.  The isoline corresponding to  
$\Delta \chi ^{2}=100$ is shown in Fig.~\ref{fig:AllContours} by the solid line.  
The acceptable fits lie inside a trough that extends from large gluino masses and
$\alpha _{s}(M_Z)=0.118$ down to $m_{\tilde{g}}\approx 0.8$~GeV and right
to $\alpha _{s}(M_Z)=0.145$.  An even narrower area corresponds to fits with 
$\chi ^{2}$ close to those in the CTEQ6M fit.  We note that $\chi ^{2}$ is 
better than in the CTEQ6M fit in a small area in which 
$m_{\tilde{g}}\lesssim 20$~GeV and $\alpha _{s}(M_{Z})\gtrsim 0.125$,
with the minimum $\Delta \chi ^{2}\approx -25$ at $m_{\tilde{g}}=8$~GeV
and $\alpha _{s}(M_Z)=0.130$.  This negative excursion in $\Delta \chi ^{2}$ 
is smaller than the tolerance $T^2$ and should therefore not be
interpreted as evidence for a light gluino.

A substantial region of $\alpha _{s}(M_Z)$ and $m_{\tilde{g}}$ is excluded
by the criterion $\Delta \chi ^{2}<100$.  For $\alpha _{s}(M_{Z})=0.118$, 
gluinos lighter than $12$~GeV are disfavored.  However, the lower bound on 
$m_{\tilde{g}}$ is relaxed to less than $10$~GeV if $\alpha _{s}(M_{Z})$ 
is increased above $0.120.$   

In Fig.~\ref{fig:AllContours}, the positions are marked of the points 
$\{\alpha _{s}(M_Z),m_{\tilde{g}}\}$ of the best fits in earlier 
PDF analyses with a 
light gluino~\cite{Roberts:1993gd,Blumlein:1994kw,Ruckl:1994bh,Li:1998vr}.%
~\footnote{The points corresponding to fits with gluino mass 
$m_{\tilde{g}}<0.7$~GeV in Refs.~\cite{Blumlein:1994kw,Ruckl:1994bh,Li:1998vr} 
are off scale and are not shown.} 
Most of these earlier solutions are excluded by the present data set, with 
the exception of the fits corresponding
to $m_{\tilde{g}}=5$ GeV and large $\alpha _{s}(M_{Z})=0.124,$ $0.129$,
and $0.134$ \cite{Roberts:1993gd,Blumlein:1994kw}.   

\begin{figure}
\begin{center}\includegraphics[  width=1.0\textwidth,
  keepaspectratio]{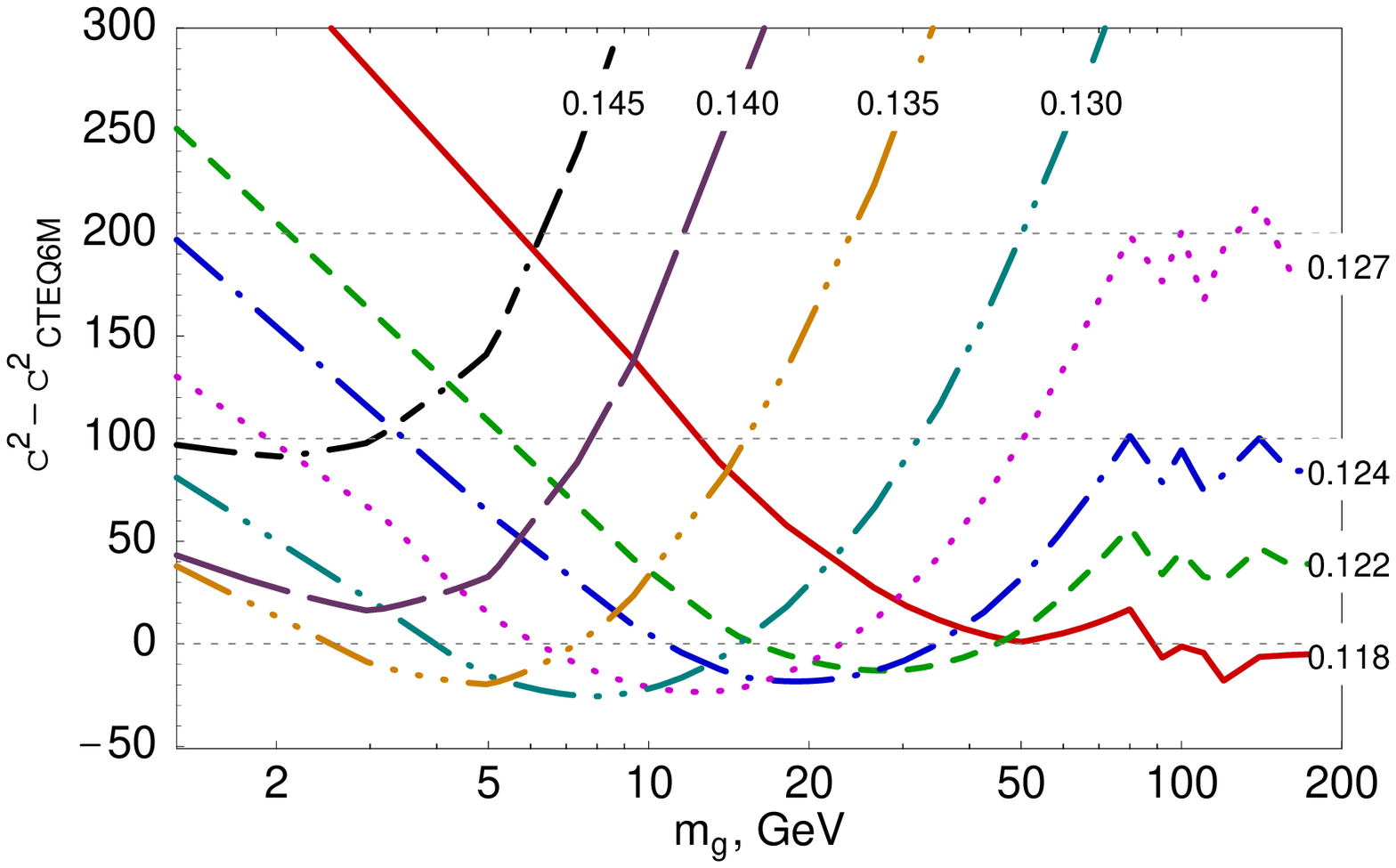}\end{center}

\caption{Dependence of $\chi ^{2}$ on $m_{\tilde{g}}$ for several fixed
values of $\alpha _{s}(M_{Z})$ (denoted by labels on each corresponding
line).   \label{fig:AMparm}}
\end{figure}

Another perspective is provided by a plot of $\chi ^{2}$ vs. $m_{\tilde{g}}$
for several fixed values of $\alpha _{s}(M_{Z})$, shown in Fig.~\ref{fig:AMparm}.
The dependence on $m_{\tilde{g}}$ is observed to be quasi-parabolic, with
a shift of the minimum of $\chi ^{2}$ to lower $m_{\tilde{g}}$ as
$\alpha _{s}(M_{Z})$ increases.  When $\alpha _{s}(M_{Z})$
is close to the current world-average value of 0.118, the fit prefers
a heavy gluino, or no gluino at all. Very light gluinos are strongly
disfavored, and the bound $m_{\tilde g} > 12$ GeV is obtained for
$\Delta \chi^2 < 100$. For $\alpha_s = 0.122$ and $0.124$, the
corresponding bounds are $m_{\tilde g} > 5$ GeV and 
$m_{\tilde g} > 3$ GeV, respectively.

Conversely, for a very large $\alpha _{s}(M_Z)$ ($>$~0.127), the pattern
is reversed, and lighter gluinos ($m_{\tilde g} < 50$ GeV) 
are preferred. In the transition region 
of $\alpha _{s}(M_Z)$ about 0.127, both very light and very heavy gluinos are 
disfavored, and a gluino mass in the range $10$ to $20$~GeV yields a 
slightly better $\chi ^{2}$ than in the CTEQ6M fit.
For a very small gluino mass of $1.3$ to $5$~GeV, the minimum in 
$\chi ^{2}$ is 
achieved for $\alpha _{s}(M_Z)$ about 0.135, while even larger values of 
$\alpha _{s}(M_Z)$ are disfavored as well (c.f. the curves for 
$\alpha _{s}=0.140$ and $0.145$).

The behavior of $\chi^2$ in Figs.~\ref{fig:AllContours} 
and~\ref{fig:AMparm} exhibits irregular structure when the gluino mass  
lies in the range $50$ to $\sim 200$~GeV.~\footnote{The irregularities 
of the contours are smoothed in Fig.~\ref{fig:AllContours}.}  These 
gluino masses lie beyond the range of sensitivity of the data sets in 
the fit, with the exception of the Tevatron jet data at jet transverse 
energies $E_T > 2m_{\tilde g}$.  Variations in $\chi^2$ at high 
gluino masses are caused by an interplay between the gluino 
mass and individual CDF and D\O{} jet data points. 
Contributions from gluinos with masses of $100 - 140$~GeV improve the 
description of very high-$E_T$ jet events, leading to a dip in $\chi^2$ 
in this region. Further discussion is found in Sec.~\ref{sec:jetsection}.  

In summary, as $\alpha _{s}(M_{Z})$ increases, the window of 
allowed gluino masses shifts from high to low values. 
Very large values of $\alpha _{s}(M_Z)(\gtrsim 0.148)$ can be ruled 
out for any gluino mass. Gluino masses between $10$ and $20$~GeV are 
allowed, as long as $\alpha_{s}(M_Z)$ is not smaller than $\sim $0.118.

\subsection{Exploration of the light gluino fits} 
\begin{figure}
\begin{center}\includegraphics[  width=1.0\textwidth,
  keepaspectratio]{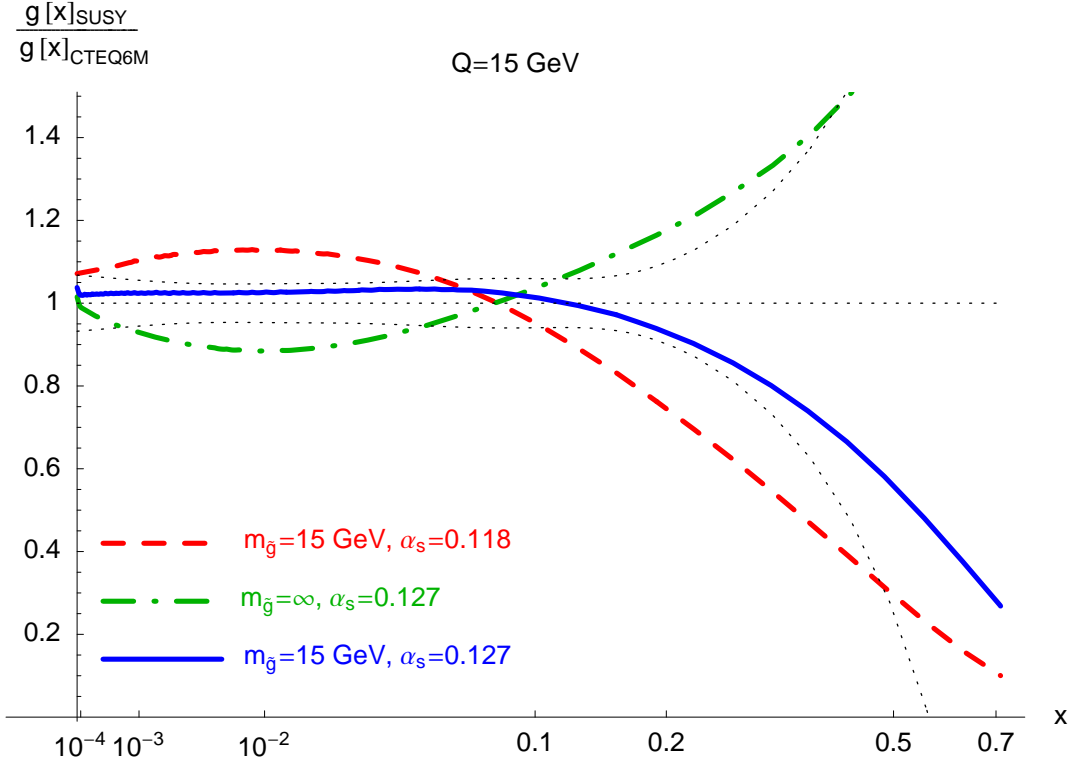}\end{center}

\caption{Illustration of the compensation in the gluon density that 
arises between an increase in $\alpha_{s}(M_Z)$ and a finite gluino 
mass.  Shown are the ratios of the gluon density in the LG solution 
divided by the CTEQ standard model 
best fit ($m_{\tilde{g}}= \infty$ and $\alpha _{s}(M_Z) = 0.118$) 
at $Q=15$~GeV.  The gluino 
mass is $m_{\tilde{g}}=15$~GeV or $m_{\tilde{g}}= \infty$, as specified 
in the figure. The CTEQ6 uncertainty band is shown by the dotted 
lines. \label{fig:gratio}}
\end{figure}

The contour plot in Fig.~\ref{fig:AllContours} indicates that excellent 
fits to the global data can be obtained with a gluino mass below the 
weak scale, $m_{\tilde{g}}\lesssim 100$~GeV, provided that $\alpha _{s}(M_Z)$
is allowed to increase above the nominal value $\alpha _{s}(M_Z) = 0.118$.  
It is instructive to examine the compensating effects of 
$\alpha _{s}(M_Z) > 0.118$ and finite gluino mass on the parton 
distribution functions themselves.  

In Fig.~\ref{fig:gratio}, the ratio shown as a dashed line provides a 
comparison of the gluon distribution $g(x,Q^{2})$ at $Q= 15$~GeV and gluino 
mass $m_{\tilde{g}} = 15$~GeV with $g(x,Q^{2})$ in the CTEQ6M fit (without 
gluinos).  The strong coupling $\alpha _{s}$ at the scale $M_{Z}$ is chosen 
to be the same as in the CTEQ6M fit, $\alpha _{s}(M_{Z})=0.118$.  The 
ratio shows that, as the gluino mass is decreased below the weak scale, 
$g(x,Q^{2})$ is depleted at large $x$ and increased at small $x$. This 
softening of the gluon distribution follows from the slower evolution of 
$\alpha _{s}(Q)$, as well as from the presence of the additional coupling 
$g\rightarrow \tilde{g}\tilde{g}$. 

For the same $\alpha _{s}(M_{Z})$, the magnitude of $\alpha _{s}(Q)$
at scales $Q < M_{Z}$ is smaller in the LG case than in the SM case 
(cf.~Fig.~\ref{fig:alphaSLG}b).  Correspondingly, PDF evolution
is slower in the LG case. To some degree, the effects of the slower
backward evolution can be compensated by selection of a larger value of  
$\alpha_{s}(M_{Z})$.  In some range of $m_{\tilde{g}}$ and 
$\alpha_{s}(M_Z)$, the effects of a smaller light gluino 
mass can be offset by a larger value of $\alpha _{s}(M_{Z})$. 
This statement is illustrated by the dot-dashed curve in 
Fig.~\ref{fig:gratio}.  The dot-dashed curve shows the ratio of 
the gluon density for $m_{\tilde g} = \infty$ and 
$\alpha _{s}(M_{Z})$ increased arbitrarily to 
$0.127$, divided by the SM CTEQ6M result ($m_{\tilde g} = \infty$ and 
$\alpha _{s}(M_{Z})=0.118$).  
The comparison shows that, when $\alpha _{s}(M_{Z})$ is increased, 
$g(x,Q^{2})$ is enhanced at large $x$ and depleted at small $x$.
The solid line in Fig.~\ref{fig:gratio} shows that, by lowering  
$m_{\tilde{g}}$ below $100$~GeV for a fixed $\alpha _{s}(M_{Z})$, we 
can approximately cancel the effect of increasing $\alpha _{s}(M_{Z})$ 
at a fixed $m_{\tilde{g}}$. The solid line lies within the band of 
uncertainty of the CTEQ6 gluon density, indicative of a fit of good 
quality.  The cancellation breaks down at very large 
$\alpha_{s}(M_Z)$. 

A similar cancellation between the effect of a small $m_{\tilde{g}}$ and 
increased $\alpha _{s}(M_{Z})$ is apparent in the singlet quark PDF. 
Consequently, a region exists at small 
$m_{\tilde{g}}$ and large $\alpha _{s}(M_{Z})$ where the resulting 
PDFs remain close to those in the CTEQ6M fit.  If $\alpha_{s}(M_Z)$ is 
allowed to float freely in the fit, one can obtain PDFs 
similar to the CTEQ6M PDFs for all values of $m_{\tilde{g}}$ above 0.8
GeV.  
For $m_{\tilde{g}} \gtrsim 150$~GeV, the PDFs are practically 
the same as in the CTEQ6M fit, indicating that the current inclusive 
hadronic data are not sensitive to such heavy particles. 

%
%

\subsection{Impact of various data sets}

To appreciate which data are the most restrictive in our fits, we 
examine the roles played in the fit by the hadronic jet data and 
other experiments.  

\subsubsection{Tevatron jet data}
\label{sec:jetsection}

The Tevatron jet data places important constraints on
$m_{\tilde{g}}$.  In the absence of the jet data, the lower limit on 
$m_{\tilde{g}}$ is weaker, with $m_{\tilde{g}}\gtrsim 5$~GeV 
at $\alpha_s(M_Z) =0.118$ if the jet data are omitted, 
but $m_{\tilde{g}}\gtrsim 12$~GeV if the jet data are included. 

\begin{figure}
\includegraphics[  width=0.49\textwidth,
  keepaspectratio]{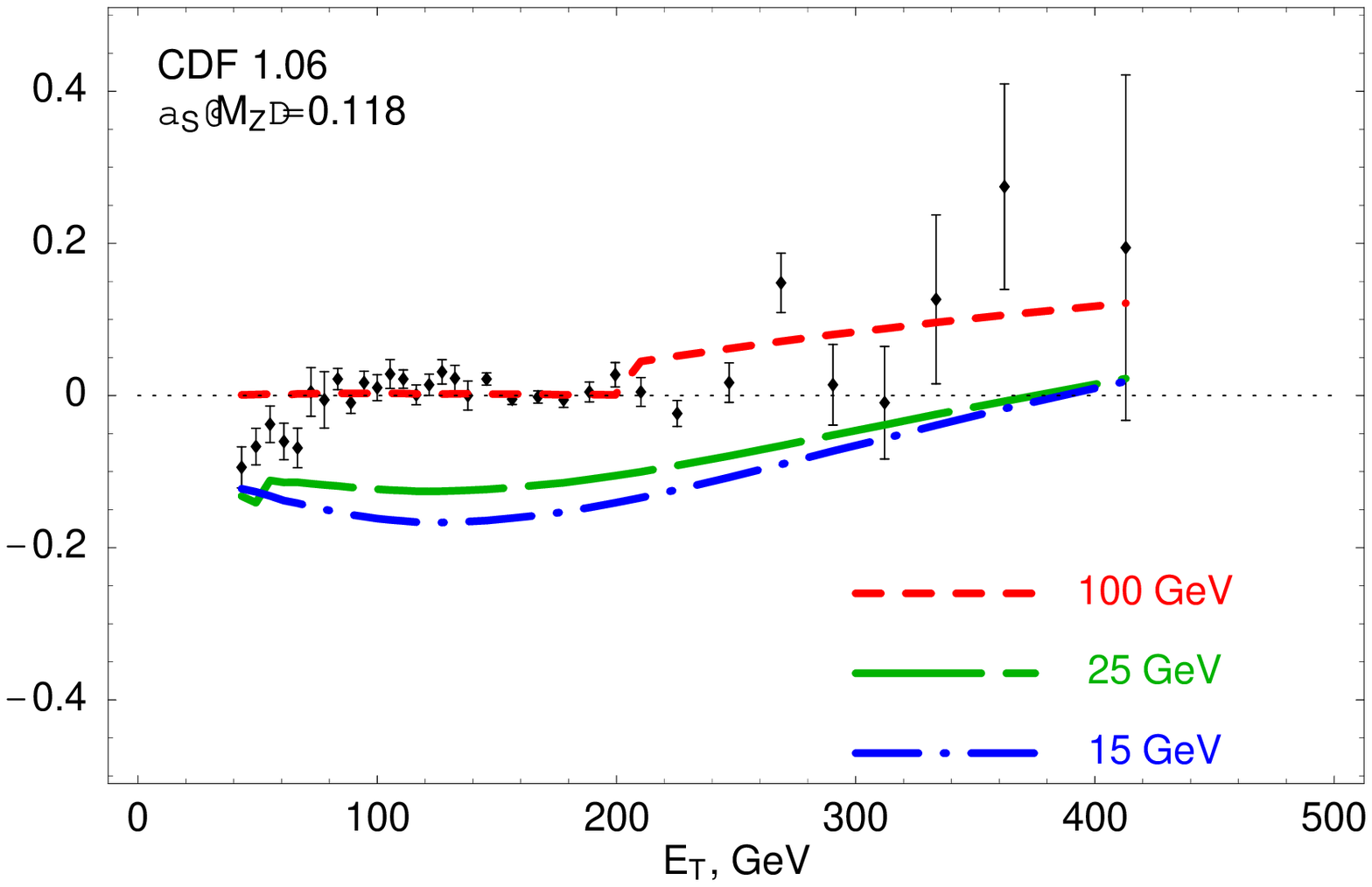} \includegraphics[  width=0.475\textwidth,
  keepaspectratio]{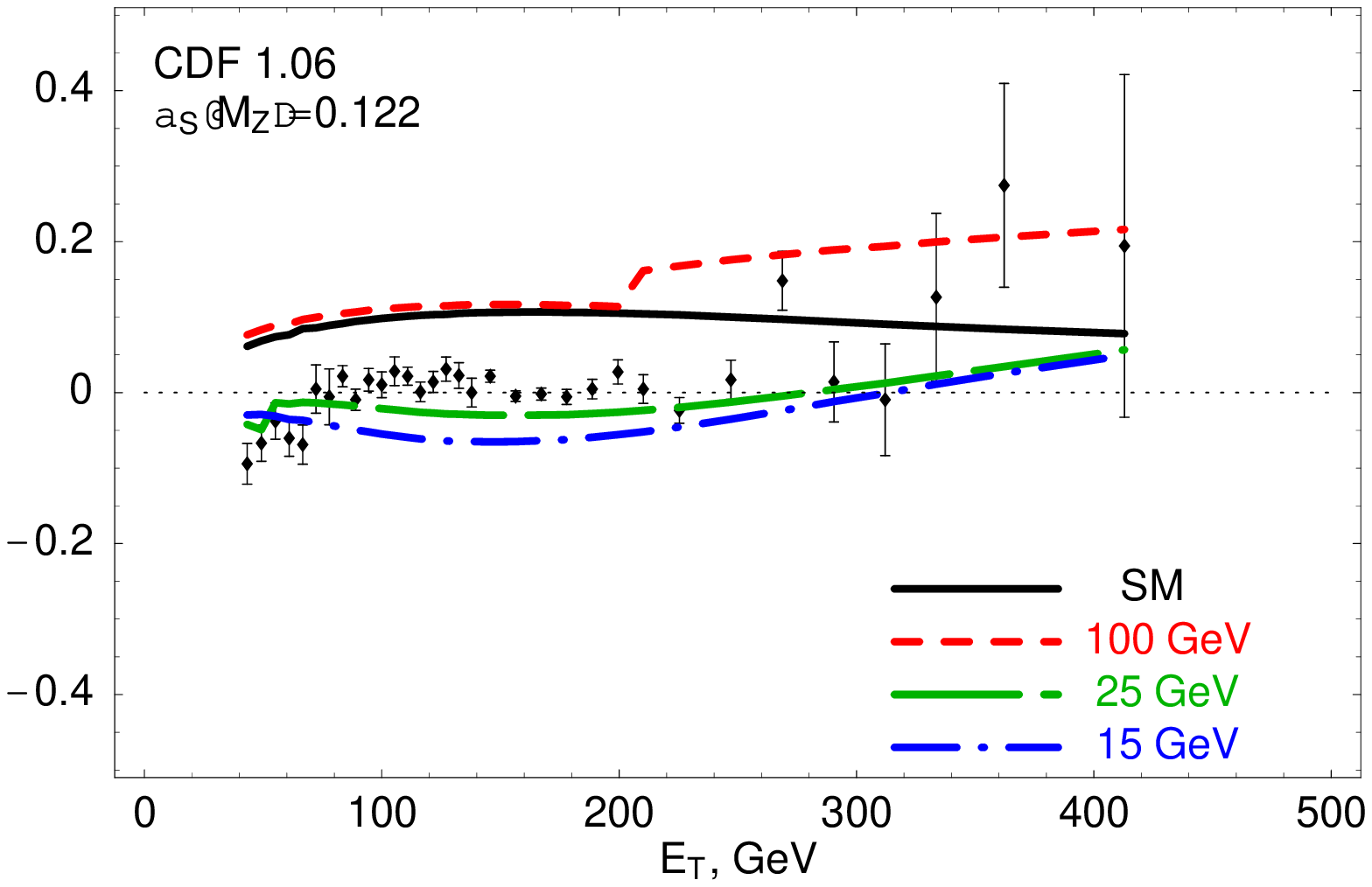} 

\includegraphics[  width=0.475\textwidth,
  keepaspectratio]{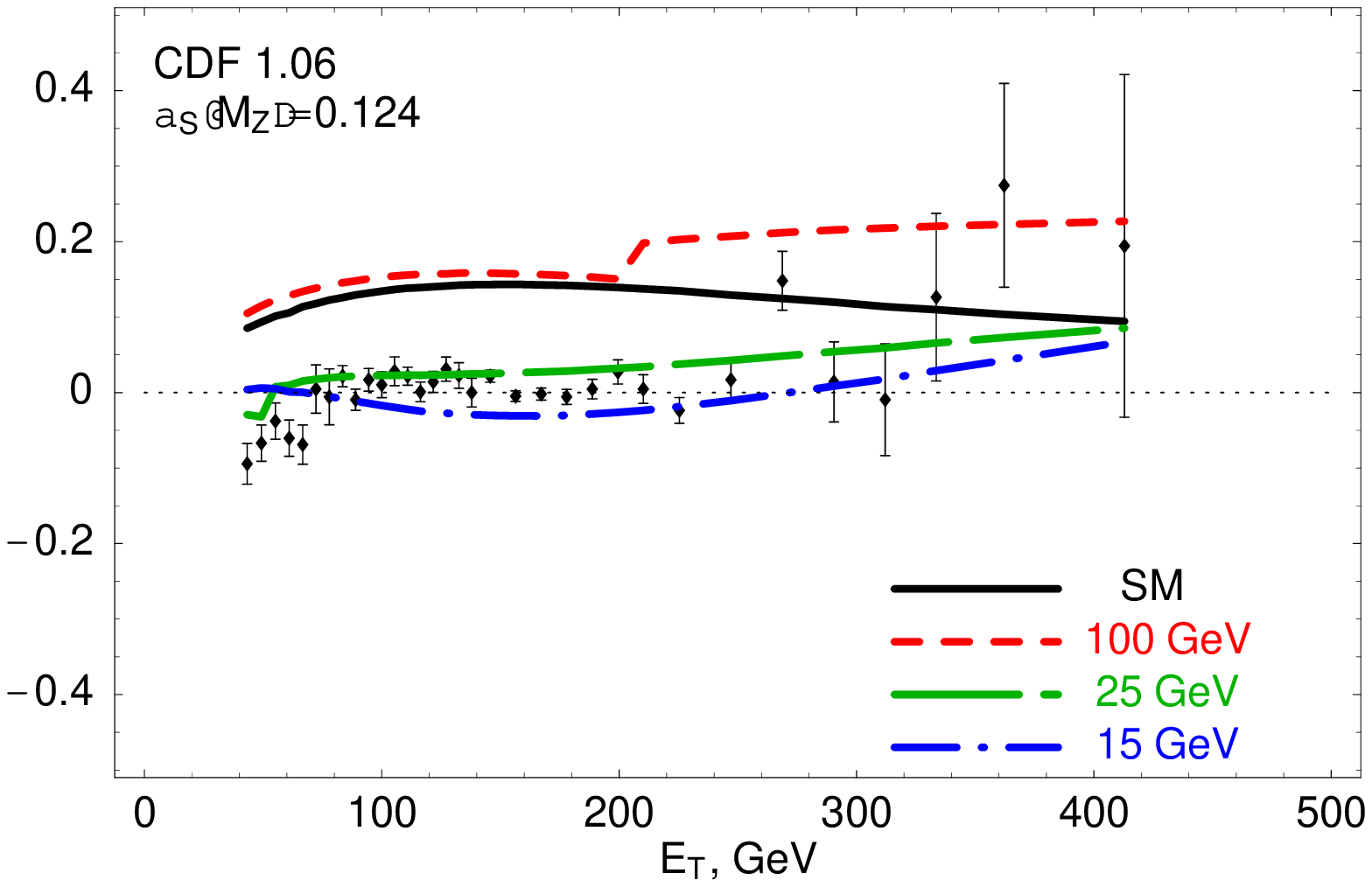}

\caption{Comparison of theoretical predictions with the CDF inclusive jet
data. The data points show the ratio (Data-CTEQ6M)/CTEQ6M,
where Data denotes the CDF measurements, and CTEQ6M is the SM prediction
based on the CTEQ6M PDFs.  The curves show the ratio (Theory-CTEQ6M)/CTEQ6M,
where the Theory curves are the calculations based on the LG PDFs, for 
$\alpha _{s}(M_{Z})= 0.118$, $0.122$, and $0.124$, and
gluino masses $m_{\tilde{g}}=\{15,25,100\}$~GeV. The solid curves correspond 
to the SM fits (effectively $m_{\tilde{g}}=\infty$) for the indicated 
choices of $\alpha _{s}(M_Z)$. The horizontal scale shows the transverse 
energy $E_{T}$ of the jet in GeV units. \label{fig:cdf}}
\end{figure}

\begin{figure}
\includegraphics[  width=1.0\textwidth,
  keepaspectratio]{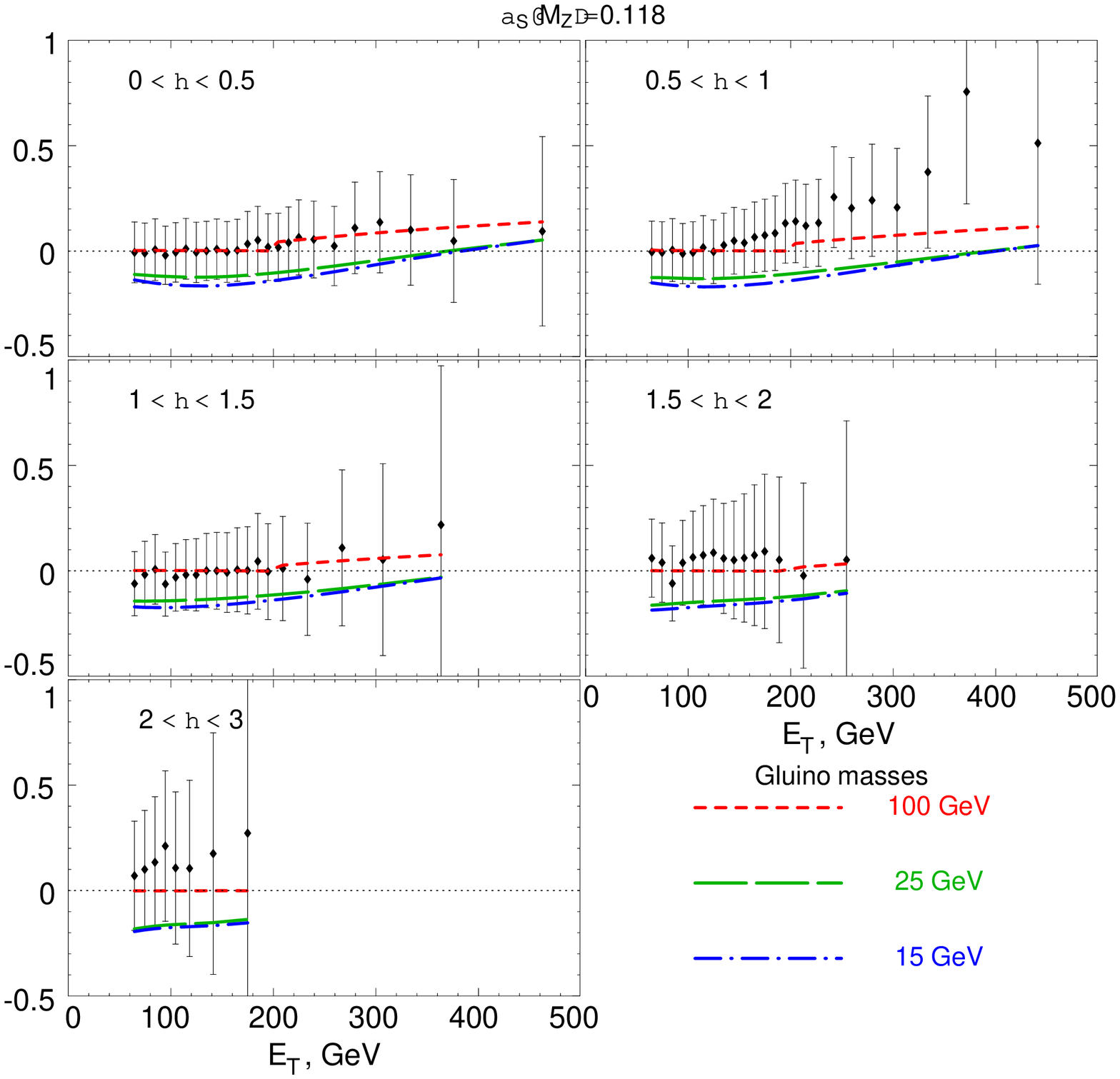}

\caption{Comparison of theoretical predictions with the D\O{} jet data for
$\alpha _{s}(M_{Z})=0.118$.  Shown are (Data-CTEQ6M)/CTEQ6M
and (Theory-CTEQ6M)/CTEQ6M, where Data are the D\O{} measurements,
and CTEQ6M and Theory are defined as in Fig.~\ref{fig:cdf}. The
data are sorted according to the ranges of the jet pseudorapidity
$\eta $.  Curves are plotted for $m_{\tilde{g}}=\{15,25,100\}$~GeV. 
\label{fig:dzero:0.118}}
\end{figure}
\begin{figure}
\includegraphics[  width=1.0\textwidth,
  keepaspectratio]{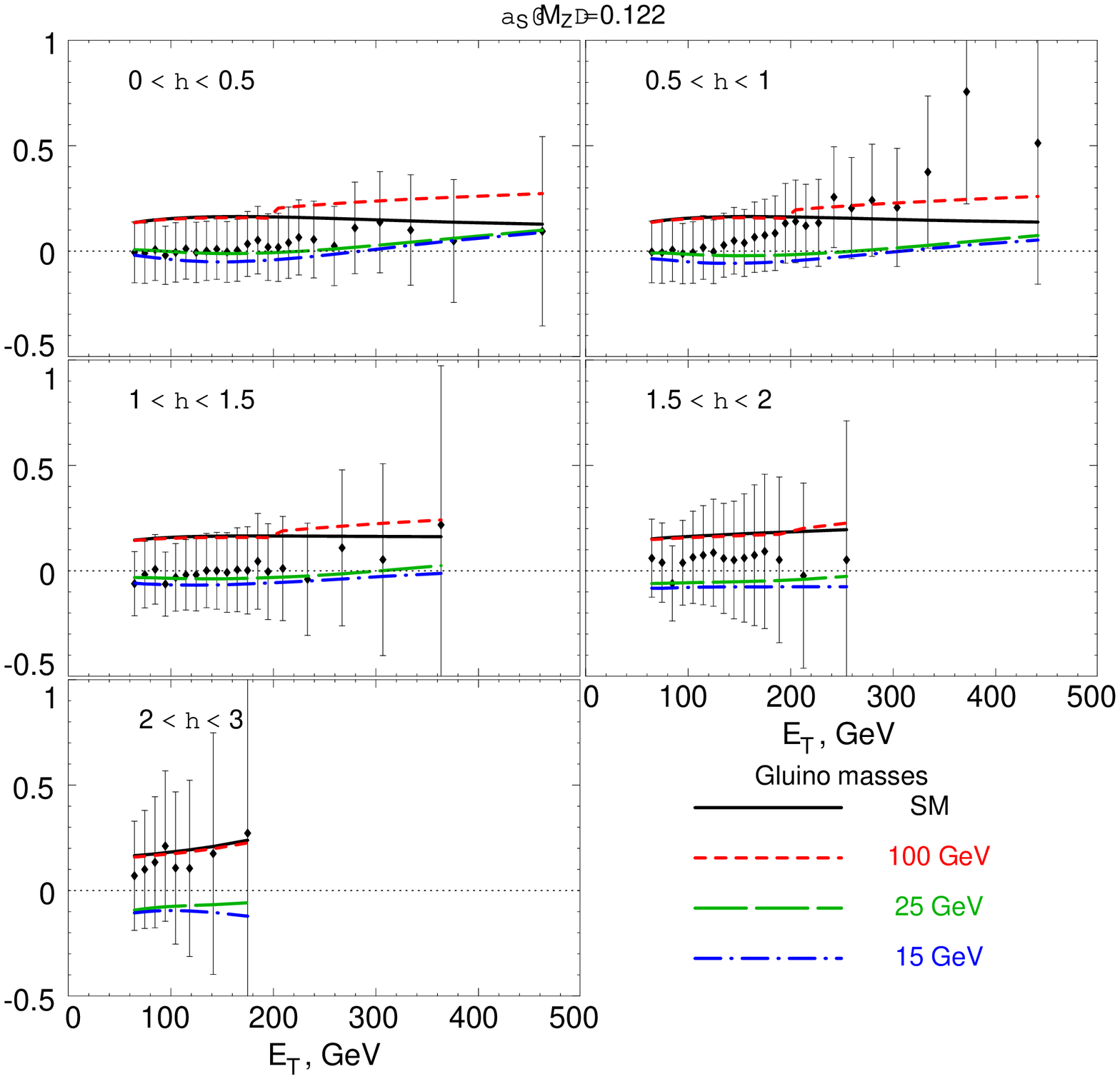}

\caption{Same as in Fig.~\ref{fig:dzero:0.118}, for 
$\alpha _{s}(M_{Z})=0.122$.
The solid curves correspond to the SM fit for 
$\alpha _{s}(M_Z)=0.122$.\label{fig:dzero:0.122}}
\end{figure}
\begin{figure}
\includegraphics[  width=1.0\textwidth,
  keepaspectratio]{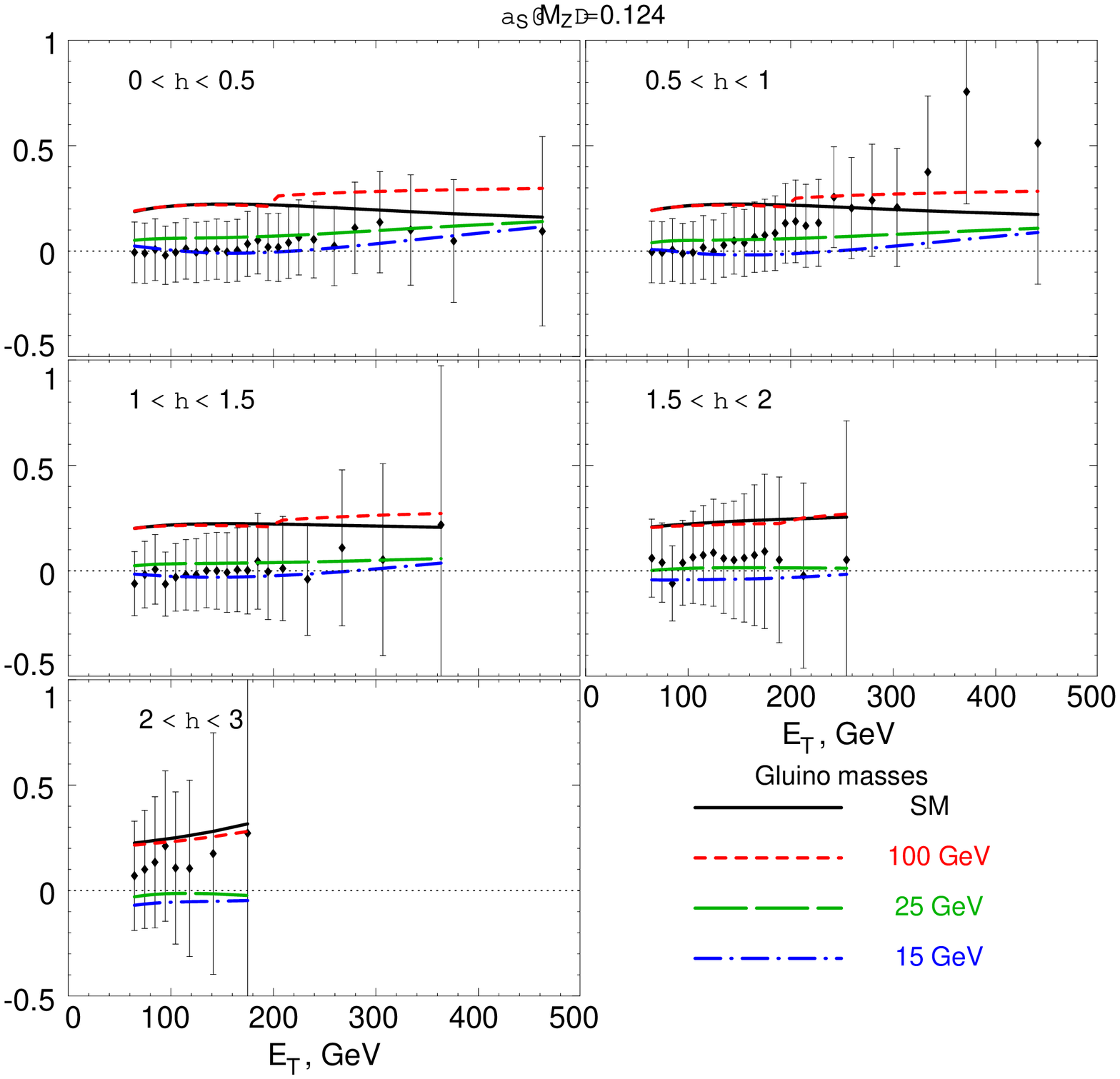}

\caption{Same as in Fig.~\ref{fig:dzero:0.118}, for $\alpha _{s}(M_{Z}) = 0.124$.
The solid curves correspond to the SM fit with 
$\alpha _{s}(M_Z) = 0.124$.\label{fig:dzero:0.124}}
\end{figure}

Comparisons between theory and the inclusive jet data from the 
CDF Collaboration~\cite{Affolder:2001fa} and the D\O{} 
Collaboration~\cite{Abbott:2000ew,Abbott:2000kp} are shown in 
Fig.~\ref{fig:cdf} and Figs.~\ref{fig:dzero:0.118}~-~\ref{fig:dzero:0.124}.  
The results are from a series of fits for fixed $\alpha _{s}(M_{Z})=0.118$, 
$0.122$, and $0.124$.  These three values of $\alpha _{s}(M_{Z})$ represent 
roughly the world-average central value and the values that are 
approximately one and 
two standard deviations larger.  The CDF data are rescaled by $1.06$, to 
account for differences in the measured luminosity used by the CDF and D\O{} 
Collaborations for the run-I data sample. The SM CTEQ6M fit with 
$\alpha _{s}(M_{Z})=0.118$ provides a good description of the data.  
For $\alpha_s(M_Z)=0.118$ and $m_{\tilde g} = 15-25$ GeV, the 
theoretical cross sections are below the data due to the depletion 
of the gluon PDF at large momentum fractions $x$ (Fig.~\ref{fig:gratio}). 
The gluon depletion can be counterbalanced in a wide range of 
$m_{\tilde g}$
by a larger value of $\alpha_s(M_Z)$. The best fits 
correspond to combinations of $\alpha_s$ and $m_{\tilde g}$ near  
the bottom of the trough in $\Delta\chi^2$ in Fig.~\ref{fig:AllContours}.
For gluinos in the range 10-25 GeV, an 
acceptable fit is possible if  $\alpha _{s}(M_Z)$ is increased 
to about $0.124$.

Contributions from gluinos increase the jet cross sections at 
$E_T > 2 m_{\tilde g}$.  A new channel for hard scattering is opened, 
and the evolution of $\alpha_s(Q)$ is slower.  For $\alpha_s(M_Z) = 0.118$,
a heavy gluino in the range $100 - 140$~GeV improves agreement of 
theory with the Tevatron jet data in the high-$E_T$ tail by augmenting the  
rate of the tightly constrained standard model contributions.
Better agreement for $m_{\tilde g}=100$~GeV~(dashed line) is visible 
in the high-$E_T$ region in Figs.~\ref{fig:cdf}(a) and~\ref{fig:dzero:0.118}.
Below the gluino threshold, the theory prediction (derived from the
fit to the data insensitive to gluino contributions) is identical to the 
CTEQ6M fit. While $\chi^2$ for the D\O{} data set
is visibly improved (cf. Fig.~\ref{fig:chi:0.118}), the reduction of 
the overall $\chi^2$ by 20 units is not statistically significant.
It will be interesting to see whether the trend in favor of contributions 
from gluinos (or other new color-charged fermions) with masses around 
$100$~GeV is preserved in the jet data from run-II at the Tevatron.

\subsubsection{Plots of the data sets vs $\chi ^{2}$}
\begin{figure}
\begin{center}\includegraphics[  width=1.0\textwidth,
  keepaspectratio]{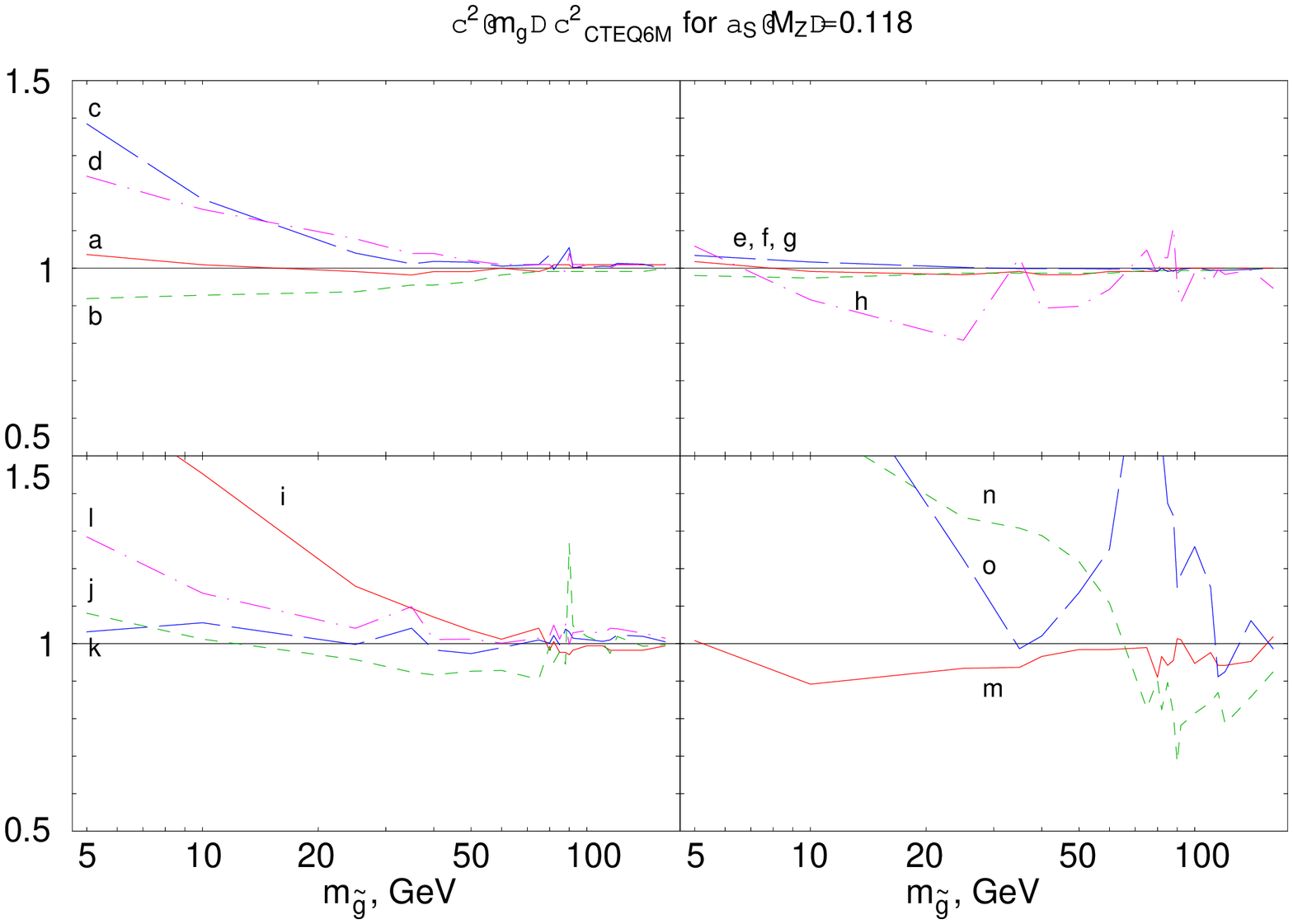} \end{center}

\caption{Ratios $\chi^{2}/\chi_{\rm {CTEQ6M}}^{2}$ for individual experiments and 
and $\alpha _{s}(M_{Z})=0.118$. 
The curves correspond to the following data sets: (a) BCDMS $F_{2}^{p}$
\cite{Benvenuti:1989rh}; (b) BCDMS $F_{2}^{d}$ \cite{Benvenuti:1990fm};
(c) H1 $F_{2}^{p}$ (1996/97) \cite{Adloff:1999ah,Adloff:2000qk};
(d) H1 $F_{2}^{p}$ (1998/99) \cite{Adloff:2000qj}; (e) ZEUS $F_{2}^{p}$
(1996/97) \cite{Chekanov:2001qu}; (f) NMC $F_{2}^{p}$ \cite{Arneodo:1997qe};
(g) NMC $F_{2}^{d}/F_{2}^{p}$\cite{Arneodo:1997qe}; (h) NMC $F_{2}^{d}/F_{2}^{p}$
\cite{Arneodo:1997kd}; (i) CCFR $F_{2}$ \cite{Seligman:1997mc};
(j) CCFR $xF_{3}$ \cite{Seligman:1997mc}; (k) E605 muon pair production
\cite{Moreno:1991sf}; 
(l) CDF lepton asymmetry \cite{Abe:1998rv}; (m) E866 muon pair production
\cite{Towell:2001nh}; (n) D\O{} jet production \cite{Abbott:2000ew,Abbott:2000kp};
and (o) CDF jet production \cite{Affolder:2001fa}. \label{fig:chi:0.118}}
\end{figure}
\begin{figure}
\begin{center}\includegraphics[  width=1.0\textwidth,
  keepaspectratio]{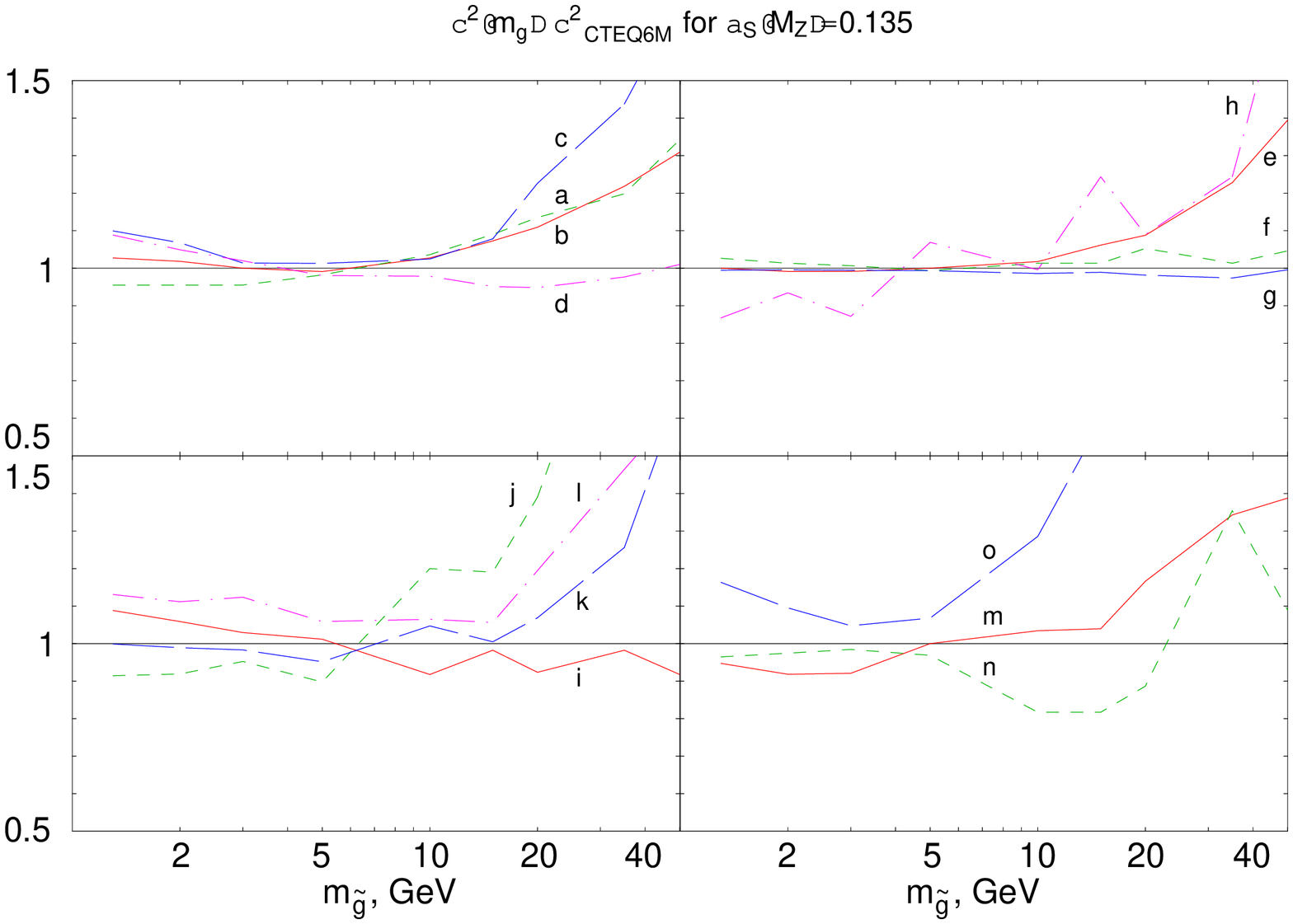} \end{center}

\caption{Same as in Fig.~\ref{fig:chi:0.118}, for $\alpha _{s}(M_{Z})=0.135$.
\label{fig:chi:0.135}}
\end{figure}

To investigate further the influence of various sets of data, we display the 
ratios $\chi^{2}/\chi_{\rm {CTEQ6M}}^{2}$ for individual experiments.   
Results for $\alpha _{s}(M_{Z})=0.118$
are shown in Fig.~\ref{fig:chi:0.118} and those for 
$\alpha _{s}(M_{Z})=0.135$ in Fig.~\ref{fig:chi:0.135}.  The choice of 
the extreme value $\alpha _{s}(M_{Z})=0.135$ is made to accentuate 
the effects we want to demonstrate.  In Fig.~\ref{fig:chi:0.118}, we observe 
that, in addition to the jet data (sets (n) and (o)), the DIS data from the H1 
Collaboration (sets (c) and (d)) and the CCFR $F_{2}$ measurement (set (i)) 
tend to drive the gluino mass to large values when $\alpha _{s}(M_{Z})=0.118$.  
On the other hand, in Fig.~\ref{fig:chi:0.135} we see that several sets of 
data are accommodated better with a light gluino if $\alpha _{s}(M_{Z})$ 
is large.

If $\alpha_s(M_Z)$ is chosen between 0.118 and 0.135, the behaviors of the values 
of $\chi^2$ for individual experiments follow a mixture of the patterns shown in 
Figs.~\ref{fig:chi:0.118} and~\ref{fig:chi:0.135}.  Several data sets 
disallow very small and very heavy gluino masses, while gluinos in the
intermediate mass range are accommodated well by the fit.

\subsection{Section summary and momentum fractions}

If $\alpha _{s}(M_Z)$ is allowed to vary freely, reasonable fits to the global 
data set are possible for essentially any gluino mass above $\sim 1$~GeV. 
However, if $\alpha _{s}(M_Z)$ is constrained from other sources, say, $\tau$ 
decay or direct measurements at $M_Z$, then a global fit to scattering data 
imposes good constraints on $m_{\tilde{g}}$.  This situation is reminiscent 
of the strong correlation between the gluon PDF and $\alpha _{s}$ observed in 
previous analyses of parton densities~\cite{Duke:1984gd,Huston:1998jj}. Similarly, 
it is not surprising that constraints 
on the gluino mass are coupled to our knowledge of the gluon PDF (constrained
by the hadronic jet data) and $\alpha _{s}(M_Z)$. 

The principal uncertainties on our quoted bounds on the gluino mass 
arise from neglect of NLO supersymmetric contributions to the PDF 
evolution (affecting the PDFs at a percent level), neglect of NLO 
SUSY-QCD corrections to jet production, and the limited precision of 
the criterion $\Delta\chi^2 < 100$ for the selection of acceptable fits.
The lower limit on the gluino mass can be relaxed if the NLO virtual-loop
SUSY-QCD corrections enhance the rate of the standard model subprocesses
in the Tevatron jet production. These uncertainties can be reduced 
in future analyses. 

We conclude this section with Table~\ref{tab:MomFrac}, in which we 
show the fraction of the proton's momentum carried by its constituents, in 
both the standard model CTEQ6M fit and in the LG fit with $m_{\tilde{g}}=15$~GeV 
and $\alpha _{s}(M_{Z})=0.122$.  

\begin{table}
\begin{tabular}{|c|c|c|}
\hline 
Parton type&
CTEQ6M&
$m_{\tilde{g}}=15$~GeV,\tabularnewline
&
&
$\alpha_{s}(M_{Z})=0.122$\tabularnewline
\hline
\hline 
$u+\bar{u}$&
25.2&
25.4\tabularnewline
\hline 
$d+\bar{d}$&
15.4&
15.5\tabularnewline
\hline 
$s+\bar{s}$&
5.3&
5.1\tabularnewline
\hline 
$c+\bar{c}$&
3.9&
3.6\tabularnewline
\hline 
$b+\bar{b}$&
2.5&
2.3\tabularnewline
\hline
\hline 
$\Sigma$&
52.3&
51.9\tabularnewline
\hline
\hline 
$g$&
47.5&
44.7\tabularnewline
\hline 
$\tilde{g}$&
0&
3.2\tabularnewline
\hline
\hline 
Total:&
100&
100\tabularnewline
\hline
\end{tabular}
\caption{Fractions of the proton's momentum carried by different 
parton species at $Q=100$~GeV in the CTEQ6M fit 
($m_{\tilde{g}} \rightarrow \infty$ and $\alpha_s(M_Z) = 0.118$) 
and in the LG fit with mass $m_{\tilde{g}}=15$~GeV and 
$\alpha_s(M_Z) = 0.122$.
\label{tab:MomFrac}
}
\end{table}

\section{Summary\label{sec:Conclusions}}

Our new analysis of the compatibility of a light gluino with inclusive 
scattering data goes beyond earlier 
studies~\cite{Roberts:1993gd,Blumlein:1994kw,Ruckl:1994bh,Li:1998vr}
in a number of aspects.  First, the current data are strikingly
more extensive than available ten years ago.  They cover both small-$x$
and large-$Q$ regions, come from a variety of experiments, and
are characterized by high precision.  The primary effects of a gluino
in the global analysis are changes in the evolution of the strong 
coupling strength and changes in the evolution of the parton distributions.
It is easier to observe these changes in a data sample with large
lever arms in $Q$ and $x$. 

Second, in contrast to previous studies, our fit includes the complete 
set of data from the CTEQ global analysis, including
the Tevatron jet production data. The major role of the jet data is
to constrain the gluon density at large values of fractional momentum $x$.   
The behavior of the gluon density at large $x$ is affected strongly by 
the presence of the gluinos in the mix.  Because they are sensitive to
gluons at large $x$, the jet data enhance the discriminating power of 
the global fit.  When the gluino mass changes, large variations
in $\chi ^{2}$ are observed, an influence that can be used to constrain 
the gluino parameter space.  In view of the strong correlations between 
the gluino mass, $\alpha_s(M_Z)$, and the gluon distribution, the constraints 
can be determined only after a consistent implementation of SUSY effects 
throughout all stages of the analysis.

The third new component in our study is a method~\cite{Pumplin:2002vw}
for quantitative interpretation of uncertainties in parton distributions.
With the help of this method, constraints on the acceptable gluino
mass can be imposed on the basis of the values of $\chi ^{2}$ obtained 
in the fits.  The main result of the paper is presented in 
Fig.~\ref{fig:AllContours}.  It shows the region of the gluino masses 
$m_{\tilde{g}}$ and QCD coupling strengths $\alpha _{s}(M_{Z})$ allowed 
by the present data.  The standard model fit prefers 
$\alpha _{s}(M_{Z})\approx 0.118$.  Gluinos with mass of a few GeV can be 
accommodated only if $\alpha _{s}(M_Z)$ is increased.  For example, gluinos 
with mass below $1$~GeV are admissible only if $\alpha _{s}(M_{Z})$ is about 
$0.130$ or larger. If one takes into account the effects of a light gluino 
on the measurement and running of $\alpha _{s}$, it is hard (if at all 
possible) to reconcile such a large value of $\alpha _{s}(M_{Z})$ with both 
low- and high-energy electroweak data. 

On the other hand, a possibility remains open for the existence of
gluinos with mass between $10$ to $20$~GeV, with a moderately
increased $\alpha _{s}$ ($\alpha _{s}(M_{Z})>0.119$).  This possibility 
is even slightly favored by the $\chi^2$ of the global fit.  A model with 
light gluinos and bottom squarks~\cite{Berger:2000mp} is not incompatible 
with the results of our PDF analysis.  Tighter constraints can be obtained 
in the near future, when new data from HERA and the Tevatron become 
available.  In particular, it will be intriguing to see whether higher 
statistics jet data at larger values of $E_T$ show indications of 
physics beyond the standard model.  The constraints we obtain depend on 
the value of $\alpha _{s}(M_{Z})$.  Uncertainties in the value of 
$\alpha _{s}(M_{Z})$ could be reduced through a consistent determination of 
$\alpha _{s}(M_{Z})$ from the CERN LEP data in a SUSY-QCD analysis, in which  
the effects of superpartners are included in the data analyses and Monte-Carlo 
simulations.  

Implementation of the gluino in our study relies only on the knowledge
of its strong interactions, which are determined uniquely by supersymmetry.
We consider only theoretically clean one-scale inclusive observables.
In this sense, our constraint $m_{\tilde{g}}>12$~GeV for $\alpha_s(M_{Z})=0.118$
should be compared to the constraint $m_{\tilde{g}}>6.3$~GeV for
the same value of $\alpha _{s}$ from the $Z$-boson width measurement
\cite{Janot:2003cr}.  
Tighter constraints on $m_{\tilde{g}}$ were
quoted by the searches for traces of gluino 
hadronization~\cite{Abdallah:2002qi,Heister:2003hc}
and a study of jet shapes~\cite{Abdallah:2002xz}. Although important,
these constraints are less general, since they involve assumptions
about the gluino lifetime or deal with several momentum scales in
the jet shape observables.  Our study demonstrates the potential
of global analysis to independently constrain new physics from 
hadron collider data.

\section*{Acknowledgments}

We thank S. Alekhin, S. Berge, Wu-ki Tung, 
and members of CTEQ Collaboration for helpful discussions.
P.M.N. and F.I.O. acknowledge the hospitality of Argonne and Brookhaven
National Laboratories and Michigan State University, where a portion
of this work was performed. Research
in the High Energy Physics Division at Argonne National Laboratory is
supported by the U.~S.~Department of Energy, Division of High Energy
Physics, under Contract W-31-109-ENG-38.
The research of P.~M.~N. and F.~I.~O. is supported 
by the U.S. Department of Energy under Grant DE-FG03-95ER40908, 
and the Lightner-Sams Foundation. 
The research of J.~P. is supported by the National Science Foundation 
under grant PHY-0354838.  


\begin{appendix}
\section*{Appendix: strong coupling strength}
\label{Appendix}

In this appendix we discuss the quantitative changes that occur in the 
evolution of the strong coupling strength $\alpha_s(Q)$ in the presence 
of a light gluino in the spectrum. 
We consider compatibility of $\tau$ decay and LEP $Z$-pole measurements of $\alpha_s$,
which constrain $\alpha_s(Q)$ at low and large momentum scales, respectively.
\begin{table}

\caption{\label{table:alphaS:Bethke} $\alpha _{s}(M_{Z})$ derived by evolution
from $\alpha _{s}(m_{\tau })=0.323\pm 0.030$~\cite{Bethke:2002rv}
for several gluino masses and in the standard model.}

\begin{center}\begin{tabular}{|c|c|c|}
\hline 
$m_{\tilde{g}},$ GeV&
$\alpha _{s}(M_{Z})$\\
\hline
\hline 
10&
$0.132\pm 0.004$\\
\hline 
25&
$0.125\pm 0.004$\\
\hline 
50&
$0.121\pm 0.004$\\
\hline 
90&
$0.118\pm 0.003$\\
\hline
\hline 
SM&
$0.118\pm 0.003$\\
\hline
\end{tabular}\end{center}
\end{table}

The measurement of $\alpha _{s}$ at LEP provides a constraint at scales of
order of the $Z$-boson mass $M_{Z}$, and the measurement of $\alpha _{s}$
in $\tau $-lepton decay provides a constraint at scales
of order of the $\tau $-lepton mass $m_{\tau }$.  If the gluino is
substantially heavier than the $\tau $-lepton, its presence in the spectrum 
cannot affect the measurement of $\alpha _{s}$ in $\tau $ decay. Therefore, 
$\alpha _{s}(Q )$ measured at the scale $Q =m_{\tau }$ is the same as 
in the standard model. Its recently quoted values are 
$\alpha _{s}(m_{\tau })=0.35\pm 0.03$~\cite{RPP:TauDecay} and 
$\alpha _{s}(m_{\tau })=0.323\pm 0.030$~\cite{Bethke:2002rv}.  On the other 
hand, the measurement of $\alpha _{s}(Q )$ at $Q =M_{Z}$ can be affected by 
light superpartners in the spectrum.  The values obtained in standard model 
analyses may have to be revised.  A recent estimate~\cite{Luo:2003uw} of 
SUSY effects on direct measurements of  $\alpha _{s}(M_{Z})$ finds values  
in the interval $(0.118-0.126)\pm 0.005$, where the variation in the central 
value arises from the uncertainty in the magnitude of SUSY-QCD corrections 
to the $Z$-boson decay width from processes such as 
$Z \to b\bar{\tilde b}{\tilde g}$ and $Z \to {\bar b}{\tilde b}{\tilde g}$.  

\begin{figure}
\begin{center}\includegraphics[  width=0.50\textwidth,
  keepaspectratio]{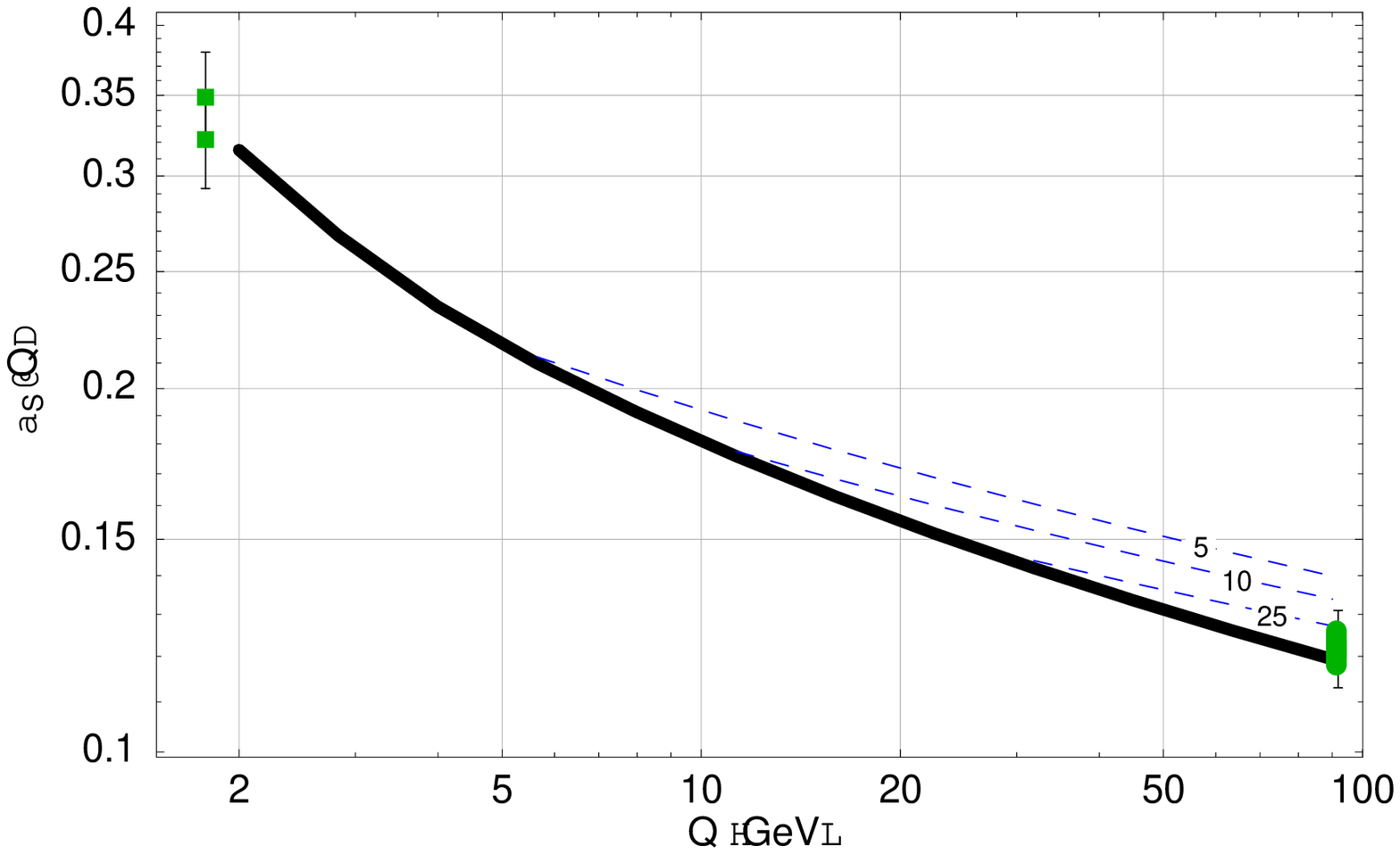}\includegraphics[  width=0.50\textwidth,
  keepaspectratio]{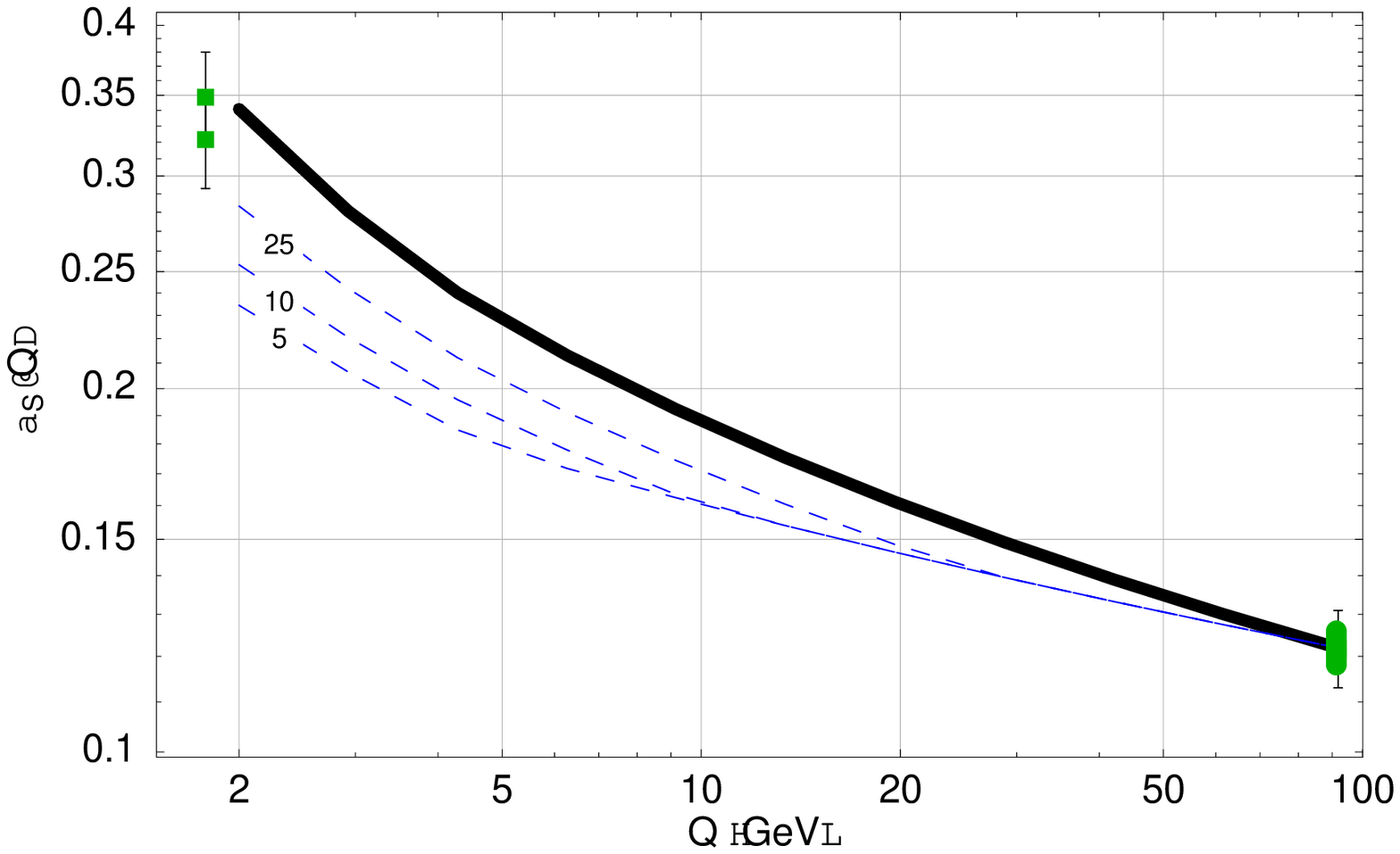}\\
(a)\hspace{7.5cm}(b)\end{center}

\caption{\label{fig:alphaSLG} Dependence of $\alpha _{s}(Q )$ on the renormalization
scale $Q $. The three data points shown are $\alpha _{s}(m_{\tau })=0.35\pm 0.03$
\cite{RPP:TauDecay}, $\alpha _{s}(m_{\tau })=0.323\pm 0.030$ \cite{Bethke:2002rv},
and $\alpha _{s}(M_{Z})=(0.118-0.126)\pm 0.005$ \cite{Luo:2003uw}.
In Fig.~\ref{fig:alphaSLG}a, the average $\alpha _{s}(m_{\tau })=0.337$
of two experimental values at $Q =m_{\tau }$ is evolved to higher
energies. In Fig.~\ref{fig:alphaSLG}b, the central value $\alpha _{s}(M_{Z})=0.122$
of the interval at $Q =M_{Z}$ is evolved to lower energies. The
thick solid line represents the Standard Model (SM) evolution in the
absence of SUSY effects. The dashed series of curves are generated
for gluino masses $m_{\tilde{g}}=5$, $10,$ and $25$ GeV (shown by the labels on 
the corresponding curves). }
\end{figure}
The $\tau $-decay and LEP values of $\alpha _{s}(Q)$ must be 
related by the renormalization group equation.
Figure~\ref{fig:alphaSLG}a shows the evolution of $\alpha _{s}(Q)$ 
measured in $\tau $ decay to the energy of order $M_{Z}$ for 
different choices of the gluino mass.  According to Eqs.~(\ref{beta0})
and (\ref{beta1}), $\alpha _{s}(Q)$ evolves more slowly in the presence 
of light gluinos.  Forward evolution of $\alpha _{s}(m_{\tau })$
in the LG case leads to a higher value at $Q =M_{Z}$ than in the
standard model. Evolution of $\alpha _{s}(m_{\tau })=0.35\pm 0.03$
results in $\alpha _{s}(M_{Z})=0.120\pm 0.003$ in the standard model
and $0.135 \pm 0.004$ for a gluino mass of $10$~GeV. 
Table~\ref{table:alphaS:Bethke}
lists the values of $\alpha _{s}(M_{Z})$
obtained by evolution from the $\tau $ decay value 
$\alpha _{s}(m_{\tau })=0.323\pm 0.030$.

Alternatively, we may evolve $\alpha _{s}$ measured in $Z$-boson 
decay backward to energies of order $m_{\tau }$ (Fig.~\ref{fig:alphaSLG}b).
The resulting $\alpha _{s}(m_{\tau })$ in the LG case is lower than 
in the standard model.  If we use the central value 
$\alpha _{s}(M_{Z})=0.122$ from the interval
$0.118-0.126$ of Ref.~\cite{Luo:2003uw} as our starting point, we obtain 
$\alpha _{s}(m_{\tau })=0.367$ in the standard model and $0.266$ for 
$m_{\tilde{g}}=10$ GeV. 

For the quoted experimental and theoretical uncertainties,
the measurements of $\alpha _{s}$ at $m_{\tau }$ and $M_{Z}$ are
incompatible at the one-standard deviation (1~$\sigma $) 
level \textit{if} the gluino mass is less than about $5.8$~GeV.  
For such gluino masses, 
the $1~\sigma$ interval of  $\alpha _{s}(M_{Z})$
obtained by evolution from $\alpha _{s}(m_\tau)$ is above 
the 1~$\sigma $ interval for the LEP measurement. 
On the other hand, the $\tau $ decay and LEP data do 
agree at the 1~$\sigma $ level for a gluino heavier than $5.8$~GeV, if
$\alpha _{s}(M_{Z})$ is at the upper end of the
theoretical uncertainty range (i.e., $\alpha _{s}(M_{Z})\sim 0.126+0.005$).
Lower values of $\alpha _{s}(M_{Z})$ increase the
lower bound on $m_{\tilde{g}}$, but gluino masses in the range $10-25$
GeV are possible within the uncertainties, as long as the central value of
$\alpha _{s}(M_{Z})$ from LEP is not less than about $0.119$.

Bounds on the gluino mass obtained from the global PDF fit depend on
the value of $\alpha_s(M_Z)$ assumed in the fit. Gluino masses lighter
than 6 GeV cannot agree simultaneously with the results of the global
fit, $\tau$ decay, and LEP $Z$ pole measurements. Gluino masses as small 
as $10$~GeV are consistent with values of $\alpha_s(Q)$ obtained 
from both $\tau $ decay and LEP data, if $\alpha_s(M_{Z})$ is increased
moderately compared to its SM world-average value of $0.118\pm 0.002$. 

\end{appendix}

\end{document}